\documentclass[manuscript]{aastex}
\usepackage{graphicx}
\usepackage{longtable}
\usepackage{lscape}
\usepackage{natbib}

\begin{document}

\title{Evolution, nucleosynthesis and yields
of low mass AGB stars at different metallicities (II): the FRUITY
database.}

  \author{S. Cristallo\altaffilmark{1,2}
    \affil{1 - Departamento de F\'isica Te\'orica y del Cosmos, Universidad de Granada, 18071 Granada,
    Spain}\affil{2 - INAF-Osservatorio Astronomico di Collurania, 64100 Teramo, Italy}}
  \and
  \author{L. Piersanti\altaffilmark{2}
    \affil{2 - INAF-Osservatorio Astronomico di Collurania, 64100 Teramo, Italy}}
  \and
  \author{O. Straniero\altaffilmark{2}
    \affil{2 - INAF-Osservatorio Astronomico di Collurania, 64100 Teramo, Italy}}
  \and
  \author{R. Gallino\altaffilmark{3,2}
    \affil{3 - Dipartimento di Fisica Generale, Universit\`a di Torino,
           10125 Torino, Italy}
    \affil{2 - INAF-Osservatorio Astronomico di Collurania, 64100 Teramo, Italy}}
  \and
  \author{I. Dom\'inguez\altaffilmark{1}
    \affil{1 - Departamento de F\'isica Te\'orica y del Cosmos, Universidad de Granada, 18071 Granada, Spain}}
    \and
  \author{C. Abia\altaffilmark{1}
    \affil{1 - Departamento de F\'isica Te\'orica y del Cosmos, Universidad de Granada, 18071 Granada, Spain}}
  \and
  \author{G. Di Rico\altaffilmark{2}
    \affil{2 - INAF-Osservatorio Astronomico di Collurania, 64100 Teramo, Italy}}
  \and
  \author{M. Quintini\altaffilmark{2}
    \affil{2 - INAF-Osservatorio Astronomico di Collurania, 64100 Teramo, Italy}}
  \and
  \author{S. Bisterzo\altaffilmark{3}
    \affil{3 - Dipartimento di Fisica Generale, Universit\`a di Torino,
           10125 Torino, Italy}}

\date{Received / Accepted}

\begin{abstract}

By using updated stellar low mass stars models, we can
systematically investigate the nucleosynthesis processes occurring
in AGB stars, when these objects experience recurrent thermal
pulses and third dredge-up episodes. In this paper we present the
database dedicated to the nucleosynthesis of AGB stars: the FRUITY
(FRANEC Repository of Updated Isotopic Tables \& Yields) database.
An interactive web-based interface allows users to freely download
the full (from H to Bi) isotopic composition, as it changes after
each third dredge-up episode and the stellar yields the models
produce. A first set of AGB models, having masses in the range 1.5
$\le M/M_{\odot} \le $ 3.0 and metallicities $1\times 10^{-3} \le Z
\le 2\times 10^{-2}$, is discussed here. For each model, a
detailed description of the physical and the chemical evolution is
provided. In particular, we illustrate the details of the {\it
s}-process and we evaluate the theoretical uncertainties due to
the parametrization adopted to model convection and mass loss. The
resulting nucleosynthesis scenario is checked by comparing the
theoretical [hs/ls] and [Pb/hs] ratios to those obtained from the
available abundance analysis of {\it s}-enhanced stars. On the
average, the variation with the metallicity of these spectroscopic
indexes is well reproduced by theoretical models, although the
predicted spread at a given metallicity is substantially smaller
than the observed one. Possible explanations for such a difference
are briefly discussed. An independent check of the third dredge-up
efficiency is provided by the C-stars luminosity function.
Consequently, theoretical C-stars luminosity functions for the
Galactic disk and the Magellanic Clouds have been derived. We
generally find a good agreement with observations.
\end{abstract}



\section{Introduction}\label{intro}

Asymptotic Giant Branch (AGB) stars are among the most efficient
polluters of the Interstellar Medium (ISM). Due to the
simultaneous action of internal nucleosynthesis (p, $\alpha$ and n
captures), convective mixing and a strong stellar wind, these
stars provide a substantial contribution to the chemical evolution
of the Universe.  Light (e.g. He, Li, F, C, N, Ne, Na, Mg) and
heavy (from Sr to Pb) elements can be produced. In particular, the
activation of neutron sources in their He-rich zone makes AGB
stars a special site for the stellar nucleosynthesis. The main
neutron source is the $^{13}$C($\alpha$,n)$^{16}$O, which is
particularly important in low mass AGB stars ($M<3 M_\odot$)
(\citealt{stra95,ga98}; for a review, see \citealt{he05} and
\citealt{stra06}). During the Thermally Pulsing AGB phase
(TP-AGB), a recurrent penetration of the convective envelope in
the H-exhausted core (Third Dredge-Up, hereinafter TDU), brings a
few protons in a region where the chemical pattern is dominated by
helium (80\%) and carbon (20\%). Then, when the temperature rises
in this region, a $^{13}$C pocket forms where the
$^{13}$C($\alpha$,n)$^{16}$O reaction takes place. Such a $^{13}$C
burning  gives rise to the so-called main component of the {\it
s}-process nucleosynthesis. It occurs in radiative conditions on a
relatively large timescale (some 10$^4$ yr), and produces a rather
low neutron density (10$^6$--10$^7$ neutrons/cm$^3$). A second
neutron source may be activated within the convective zone
generated by a Thermal Pulse (TP), the
$^{22}$Ne($\alpha$,n)$^{25}$Mg
\citep{came60,truran,cosner,ibre83}. In this case, the timescale
of the neutron capture nucleosynthesis is definitely shorter (some
yr) and the average neutron density is higher
({10$^{9}$--10$^{11}$ neutrons/cm$^3$). This second neutron burst
is only marginally activated in low mass stars of solar-like
metallicity, while it is expected to be more efficient in more
massive AGB stars and at the low metallicities typically found in
the Galactic halo or in dwarf spheroidal galaxies
\citep{stra00,kakka06,cri09a}.

From \citet{la95}, the theoretical scenario here recalled has been
largely confirmed by the abundance analysis of AGB stars
undergoing TDUs. Theoretical stellar models are essential to
interpret the continuously growing number of spectroscopic data
collected with the latest generation of modern telescopes, from
very metal-poor C-rich stars  (see \citealt{bi10} and references
therein) to the Galactic and extragalactic C (N-type) stars
\citep{ab01,ab02,abia08,ab10,dela06,zamora09}, Ba-stars
\citep{ab06,smilia,liu}, post-AGB stars \citep{vanwi,reddy,rey04},
planetary nebulae \citep{sharpee,stedi} and born-again AGB stars
(PG 1159 stars) \citep{wehe,werner}. Similarly, the growing number
of measured isotopic ratios in pre-solar SiC grains
\citep{hoppe,lu99,zinner,barz,davis} require specific calculations
to interpret them. Nucleosynthesis predictions are also needed to
built up chemical evolution models (see e.g.
\citealt{tra99,romano,koba}). All these studies require
updated theoretical predictions of the nucleosynthesis occurring
in AGB stars belonging to very different stellar populations, with
different masses and initial compositions. For these reasons, we
decided to create a complete and homogeneous database of AGB
nucleosynthesis predictions and yields.

During the last 20 years, our working group has produced several
improvements in the physical description of AGB stellar interiors
and has developed the numerical algorithms needed to follow the
complex interplay between nuclear burning, convection and mass
loss. All these efforts converged in a recent paper
(\citealt{cri09a}, hereafter Paper I), where stellar models
calculations of  2 $M_\odot$ AGB stars of various metallicities
were presented. These calculations were based on a full nuclear
network, from H to Bi, including 700 isotopes and about 1000
nuclear processes among strong and weak interactions. In this
paper, we present the FRUITY database (FRANEC Repository of
Updated Isotopic Tables \& Yields), which is available on the web
pages of the Teramo Observatory
(INAF)\footnote{http://www.oa-teramo.inaf.it/fruity}. This
database has been organized under a relational model through the
MySQL Database Management System. This software links input data
to logical indexes, optimizing their arrangement and speeding up
the response time to the user query. Its web interface has been
developed through a set of Perl{\footnote{Perl is a
general-purpose, object-oriented programming language.} scripts,
which allow to submit the query strings resulting from filling out
appropriate fields to the managing system. It contains our
predictions for the surface composition of AGB stars undergoing
TDU episodes and the stellar yields they produce. Tables for AGB
models having initial masses 1.5 $\le M/M_{\odot} \le$ 3.0 and
$1\times 10^{-3}\le Z \le 2\times 10^{-2}$ are available. FRUITY
will be expanded soon by including AGB models with larger initial
mass and/or lower $Z$.

In Sections \ref{code} and \ref{nucleo} of the present paper we
describe the stellar models and the related nucleosynthesis
results. In Section \ref{uncer} we address the main uncertainties
affecting our models while comparisons with available photometric
and spectroscopic data are discussed in Section \ref{obs}.
Conclusions are drawn in Section \ref{concl}.

\section{The FRANEC code}\label{code}

The stellar models of the FRUITY database have been obtained by
means of the FRANEC code
\citep[Frascati~RAphson-Newton~Evolutionary~Code~-][]{chi98}. The
most important upgrades needed to follow the AGB evolution and
nucleosynthesis have been extensively illustrated in
\citet{stra06}, \citet{cri07} and in Paper I.

We have computed 28 evolutionary models, for different combination
of masses (1.5, 2.0, 2.5 and 3.0 $M_\odot$) and metallicities
($Z=1.0\times 10^{-3}$ with $Y$=0.245; $Z=3.0\times 10^{-3}$ and
$Z=6.0\times 10^{-3}$ with $Y$=0.260; $Z=8.0\times 10^{-3}$ and
$Z=1.0\times 10^{-2}$ with $Y$=0.265; $Z=1.4\times 10^{-2} \equiv
Z_\odot$ and $Z=2.0\times 10^{-2}$ with $Y$=0.269).  In the
present paper only scaled solar compositions are considered, whose
relative distribution has been derived according to the solar
compilation by \citet{lo03}.

In addition to the commonly adopted opacity tables calculated for
a scaled solar composition, we use C-enhanced opacity tables, as
described in  \citet{cri07}, to take into account the effects of
TDU episodes. The most relevant atomic and molecular species have
been considered. This improved radiative opacity substantially
affects the stellar radius and the mass loss rate. This in turn,
shortens the AGB lifetime hence reducing the amount of material
returned to the ISM \citep[see~also][]{marigo,weiss,vema09}. Our
code interpolates between opacity tables whose C-enhancements
($f_i$) depend on the metallicity \citep{cri07,lear}. The $f_i$
parameter is defined as the ratio between the $^{12}$C mass
fraction and the corresponding solar value scaled to the initial
$Z$ \footnote{Note that $f_i=1.8$ corresponds to C/O$\sim$1}.
Higher values of $f_i$ are needed to follow the evolution of low
$Z$ models up to the AGB tip, due to the lower initial carbon mass
fraction and the more efficient third dredge-up.

Our computations start from a proto-star, namely a model in
hydrostatic equilibrium, chemically homogeneous, fully convective
and cooler than the threshold for any nuclear burning. We follow
the evolution from the pre-main sequence up to the end of the AGB
phase, passing through the core He-flash. No core overshoot has
been adopted during core H-burning. Semi-convection was applied
during core He-burning (see also Section \ref{modteo}). We stop
the computations when, as a consequence of the mass loss, the
H-rich envelope has been reduced to the minimum mass for the
occurrence of the TDU\footnote{The minimum envelope mass for TDU
occurrence is $M_{\rm ENV}\sim (0.4\div 0.5) M_{\odot}$ at large
metallicities and $M_{\rm ENV}\sim (0.3\div 0.4) M_{\odot}$ at low
metallicities.} (see \citealt{stra03a}). In Table \ref{tab2} we
summarize the main properties of the evolutionary sequences prior
to the thermally pulsing AGB phase. For each metallicity and
initial He content we report from left to right: the initial mass
(solar units), the core H-burning lifetime (in Myr), the maximum
size of the convective core developed during the main sequence (in
solar units), the He surface mass fraction after the occurrence of
the First Dredge-Up (FDU), the luminosity at the RGB tip
(logarithm of $L$ in solar unit), the mass of the H-exhausted core
at the beginning of core-He burning ($M_{\rm He}^1$, in solar
units), the core He-burning lifetime (in Myr), the mass of the
H-exhausted core at the end of core He-burning ($M_{\rm He}^2$, in
solar units) and at the first TP ($M_{\rm He}^3$, in solar
units)\footnote{We define the first TP when the luminosity
generated by 3$\alpha$ reactions exceeds 10$^6 L_\odot$ for the
first time.}. When comparing these models with those already
published in \citet{dom99} (which where obtained with an older
release of the FRANEC code) variations typically lower than 10\%
are found. These variations are due to the many upgrades of the
input physics and to the differences in the initial composition
(mainly the He content). For a more complete description of the
pre-AGB evolution and the relevant theoretical uncertainties, the
reader may refer to \citet{dom99}.

\subsection{The Thermally Pulsing AGB phase}\label{agb}

In Tables \ref{tab3}, \ref{tab4}, \ref{tab5} we report quantities
characterizing the evolutionary sequences during TP-AGB phase for
three key metallicities: $Z$=0.02, $Z$=0.006 and
$Z$=0.001~\footnote{A pdf file with Tables corresponding to all
the computed metallicities is available on the web page of the
database. Note that some of the models with $M=2 M_\odot$ have
already been presented in Paper I but, for a better understanding,
have also been reported in this paper.}. For each model, we report
(left to right): the progressive TP number ($n_{\rm
TP}$)\footnote{Negative numbers correspond to TPs not followed by
TDU.}, the total mass ($M_{\rm Tot}$), the mass of the H-exhausted
core ($M_{\rm H}$)\footnote{At the onset of the thermal pulse.},
the mass of the H-depleted material dredged-up by the TDU episode
which follows a TP($\Delta M_{\rm TDU}$), the maximum mass of the
convective zone generated by the TP ($\Delta M_{\rm CZ}$), the
growth of the H-exhausted core during the interpulse period
preceding the current TP($\Delta M_{\rm H}$), the overlap factor
{\it r}\footnote{Namely, the overlap, in mass, between the
convective zones generated by the present and the previous TP.},
the $\lambda$ factor\footnote{Namely, the ratio between $\Delta
M_{\rm TDU}$ and $\Delta M_{\rm H}$.}, the duration of the
interpulse period preceding the current TP ($\Delta t_{ip}$), the
maximum temperature attained at the bottom of the convective zone
generated by the TP ($T_{\rm MAX}$), the surface metallicity
attained after the TDU ($Z_{surf}$), the corresponding C/O ratio
and the average number of neutrons captured per initial iron seed
nucleus during the radiative $^{13}$C pocket burning ($n_c$).
\begin{figure*}[tpb]
\centering
\includegraphics[width=\textwidth]{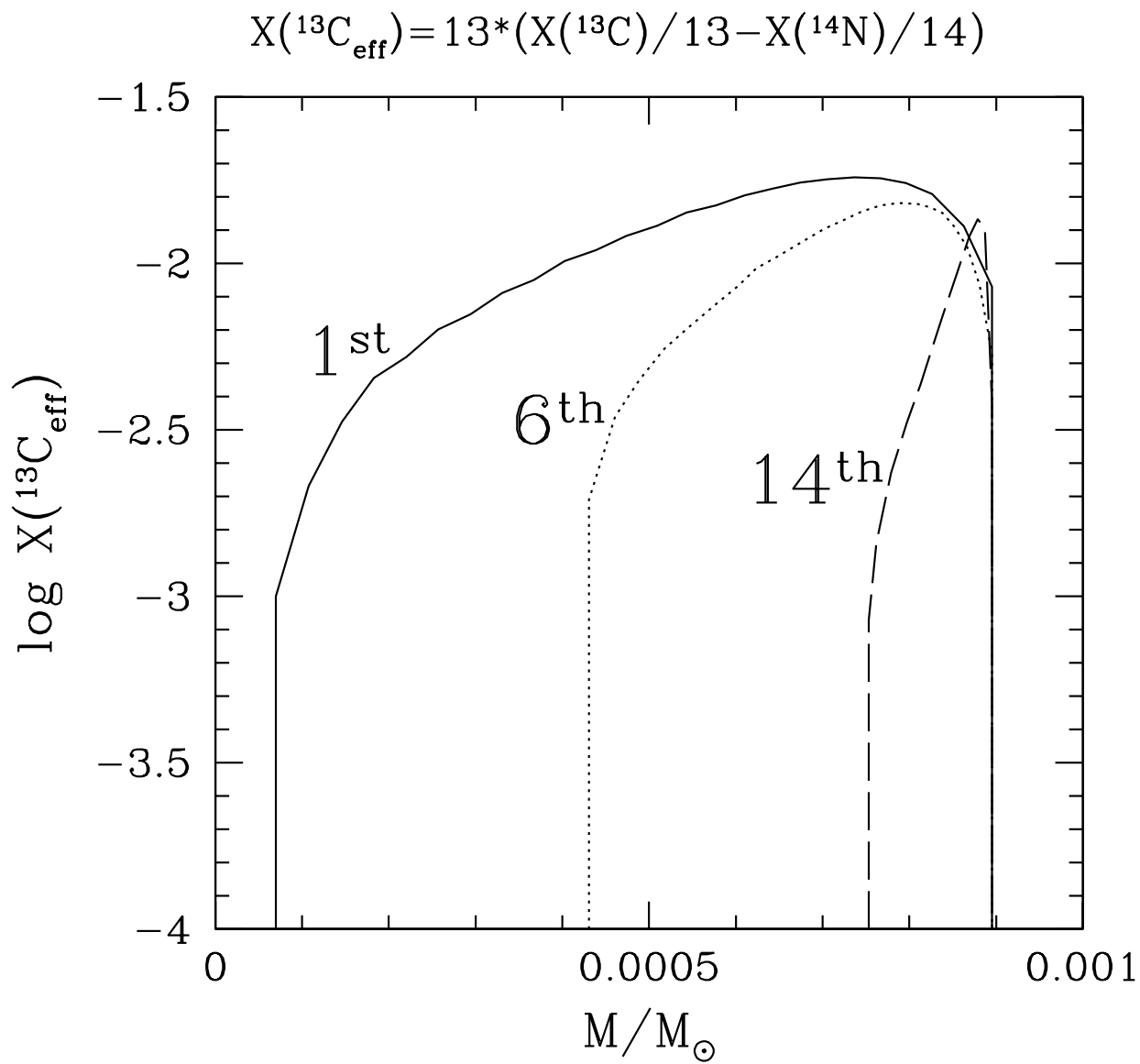}
\caption{Mass extension of the {\it effective} $^{13}$C in the
$1^{st}$, $6^{th}$ and $14^{th}$ (last) pocket of the 2 $M_\odot$
model with $Z=6\times 10^{-3}$. The pockets have been shifted in
mass in order to superimpose their external borders. The 0 point
of the abscissa is arbitrary.} \label{fig1}
\end{figure*}
Lastly, we report the total mass of the material cumulatively
dredged-up ($M_{\rm TDU}^{\rm tot}$) and the duration of the
C-rich phase of the model ($\tau_C$)\footnote{That is the time
spent of the AGB with a surface C/O$>$1.} for each model.

The main properties of the TP-AGB phase of 2 $M_\odot$ stars as a
function of the metallicity have been extensively presented in
Paper I. In this paper we explore in greater detail the evolution
of low mass stars by discussing how the physical conditions
favoring the AGB nucleosynthesis scenario change as a function of
the initial mass. One of the most important quantities driving the
nucleosynthesis in low mass AGB stars is the amount of $^{13}$C
available in the pockets, which form in the transition layer
between the H-rich envelope and the H-depleted core after each
TDU. In our models, the formation of the $^{13}$C-pocket is a
by-product of the introduction of an exponentially decaying
profile of convective velocities at the base of the envelope (see
Formula 1 in Paper I). This has been done in order to stabilize
this convective boundary during TDU episodes (see Paper I). Such a
discontinuity originates from the opacity difference between the
H-rich envelope and the He-rich intershell (i.e. the region
between the He-shell and the H-shell, hereafter He-intershell).
Even if this additional mixing always works during the TP-AGB
phase, the deriving opacity-induced extra-mixed zone is
appreciable only during TDU episodes (and thus negligible during
the interpulse phases). At the beginning of the interpulse, the
few protons in this transition layer are captured by $^{12}$C
nuclei, producing $^{13}$C (via the
$^{12}$C(p,$\gamma$)$^{13}$N($\beta^+\nu$)$^{13}$C nuclear chain).
A further proton capture may eventually lead to the production of
$^{14}$N. Consequently, after each TDU episode, a $^{13}$C pocket
forms, partially overlapped to a $^{14}$N pocket. Note that where
the abundance of $^{22}$Ne is comparable to the $^{12}$C
abundance, a third small $^{23}$Na pocket forms (see Figure 13 of
Paper I). Due to the resonant $^{14}$N(n,p)$^{14}$C neutron
capture cross section, $^{14}$N is a very efficient neutron
poison. However, the released protons give rise to a partial
neutrons recycling through the following chain:
$^{12}$C(p,$\gamma$)$^{13}$N($\beta^+$)$^{13}$C($\alpha$,n)$^{16}$O.
Since our models are calculated with a full nuclear network, the
combined effect of poisons and neutron recycling on the on-going
{\it s}-process nucleosynthesis is properly taken into account. To
illustrate the properties of our $^{13}$C pockets that follow, in
the following we use the {\it effective} $^{13}$C fraction ({\it
i.e.} the difference between the number fractions of $^{13}$C and
$^{14}$N in the pocket), rather than the bare $^{13}$C amount (see
the Formula in Fig. \ref{fig1}). Note that such a quantity is
related to the total number of neutrons available for the
subsequent {\it s}-process nucleosynthesis, which takes place when
the temperature in the pocket becomes high enough for the
activation of the $^{13}$C($\alpha$,n)$^{16}$O reaction. As
already outlined in Paper I, the pockets shrink in mass ascending
the AGB, following the compression of the whole He-intershell
region. As a typical example, in Fig. \ref{fig1} we plot the {\it
effective} $^{13}$C characterizing the $1^{st}$, $6^{th}$ and
$14^{th}$ (last) pocket of the 2 $M_\odot$ model with $Z=6\times
10^{-3}$. The pockets have been shifted in mass in order to
superimpose their external borders; the 0 point of the abscissa is
arbitrary. The extension of the pockets decreases with the TDU
number, the first being the largest one ($\Delta M\sim $9$ \times
$10$^{-4}$ $M_\odot$, about 6 times greater than the last one).\\
\begin{figure*}[tpb] \centering
\includegraphics[width=\textwidth]{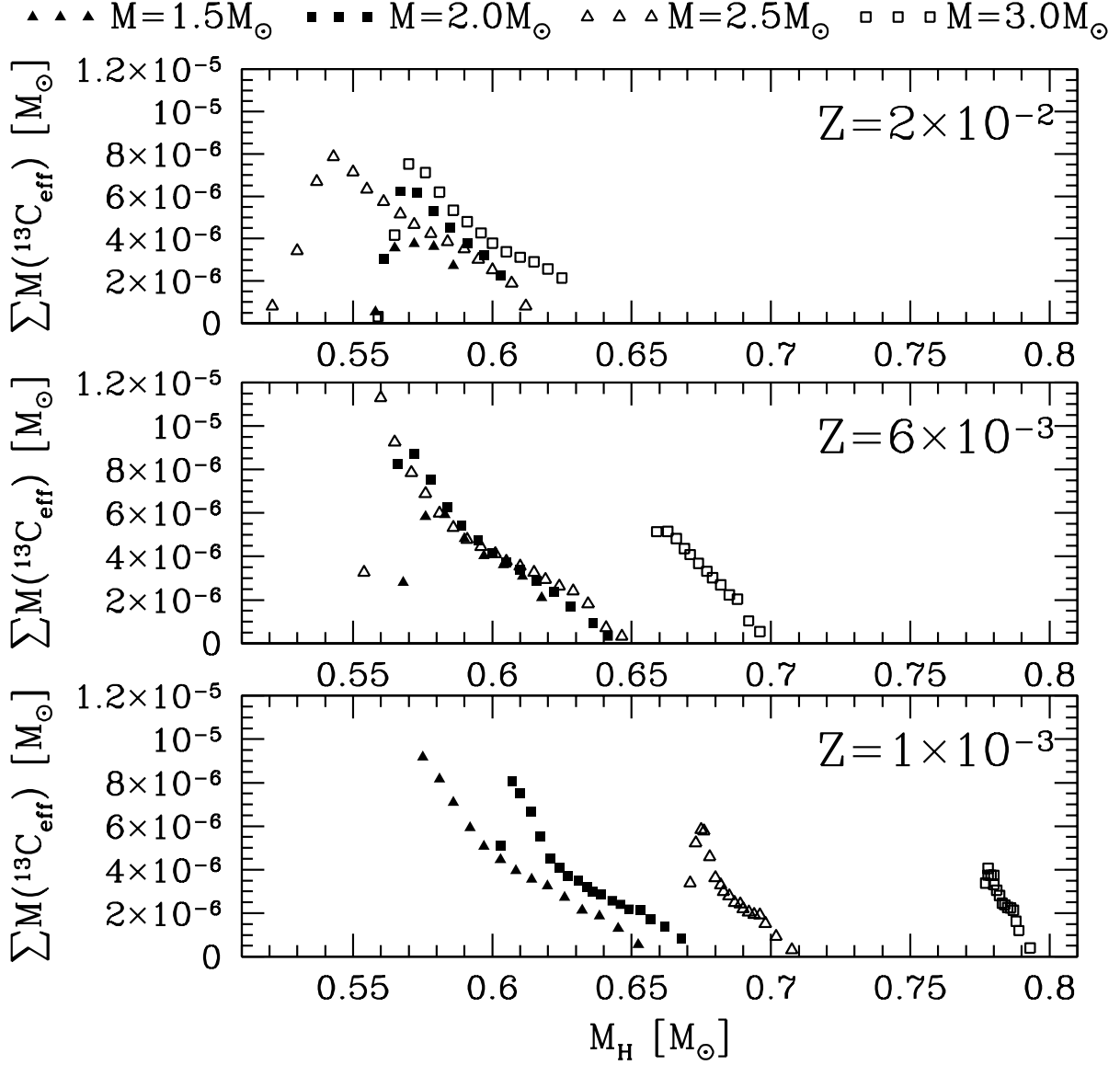}
\caption{The mass of {\it effective} $^{13}$C in the pockets
formed after each TDU episode, $\Sigma M(^{13}{\rm C}_{\rm eff})$,
versus the core mass. Symbols correspond to different initial
stellar masses, as labelled in the figure. The top panel refers to
models with $Z=0.02$, the middle panel with $Z=0.006$ and the
lower panel with $Z=0.001$.} \label{fig2}
\end{figure*}
In Fig. \ref{fig2} we illustrate the variation of the total mass
of {\it effective} $^{13}$C in each pocket\footnote{That is the
$^{13}$C mass integrated over the pocket region.} as a function of
the core mass along the evolutionary sequences corresponding to
the various initial masses. We concentrate on the selected
metallicities of Table \ref{tab2}, \ref{tab3} and \ref{tab4}
($Z=2\times 10^{-2}$, $Z=6\times 10^{-3}$ and $Z=1\times 10^{-3}$,
respectively). Each point corresponds to a single $^{13}$C pocket
and, in turn, to a single {\it s}-process nucleosynthesis episode.
In this context, the relative contribution of a single {\it
s}-process episode to the overall {\it s}-process nucleosynthesis
is proportional to the quantity reported on the $y$-axis of this
plot. As extensively discussed in Paper I, the first few pockets
dominate (qualitatively and quantitatively) the overall
nucleosynthesis of the elements during an AGB evolutionary
sequence. In practice, the composition in the He-intershell is
freezed after just 4 or 5 thermal pulses followed by TDU episodes,
being only marginally modified by the late $^{13}$C burnings. In
Paper I we outlined that, in the 2 $M_\odot$ models, the first
$^{13}$C pocket is only marginally consumed before the onset of
the following TP and thus some $^{13}$C is engulfed in the
convective shell. We confirm this finding also for the other
masses. This phenomenon is particularly strong at large
metallicities ($Z\ge 1\times 10^{-2}$), where up to 80\% of
$^{13}$C in the first pocket burns convectively. Moreover, for the
more massive and metal-rich models, a non negligible $^{13}$C
percentage ($<$30\%) in the pocket formed after the 2$^{\rm nd}$
TDU is also ingested in the TP. Lowering the metallicity, the
amount of ingested $^{13}$C progressively decreases, being
completely burnt in radiative conditions at $Z=1\times 10^{-3}$.
The convective $^{13}$C burning has only minor consequences on the
chemical evolution of the models (see Section \ref{nucleo}). Note
that even for the first two pockets a non negligible amount of
$^{13}$C (at least 20\%) burns radiatively during the interpulse
phase (the exact value depending on the mass and the metallicity).
Since the first two pockets are the largest ones, they provide an
important contribution to the $s$-process nucleosynthesis even if
the amount of $^{13}$C burnt in radiative conditions is reduced
with respect to the following pockets. In Fig. \ref{fig2} we take
into account the partial ingestion of the first pockets by
subtracting the engulfed $^{13}$C in the derivation of the {\it
effective} $^{13}$C. Changing the mass and the metallicity of the
model, a certain spread in the mass of the {\it effective}
$^{13}$C is found. Nonetheless, this occurrence does not
necessarily imply an equivalent spread in the final surface
s-process distributions, which basically depend on the neutrons
per iron seed in the pocket (see Section \ref{nucleo}). In this
connection, an interesting quantity is the average number of
neutrons per seeds in each pocket, as reported in the last column
of Tables \ref{tab3}, \ref{tab4}, \ref{tab5}. Recalling that the
first few pockets dominate the overall {\it s}-process
nucleosynthesis, we see that the neutrons captured per iron seed
substantially increases by reducing the metallicity, while, for a
fixed $Z$, it is almost insensitive to a change in the initial mass.\\
\begin{figure*}[tpb]
\centering
\includegraphics[width=\textwidth]{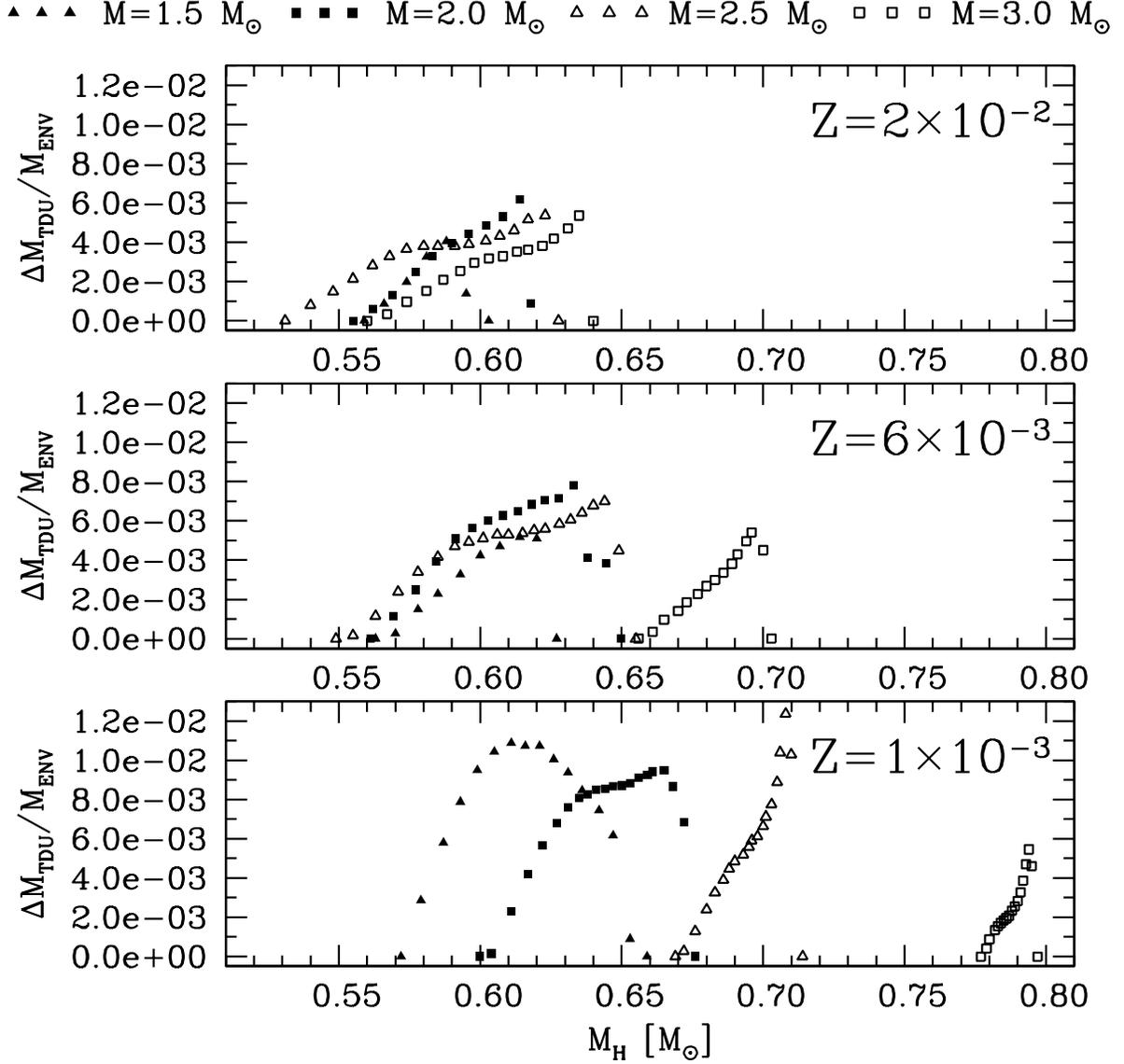}
\caption{As in Fig. \ref{fig2}, but for the ratio between the mass
of H-depleted dredged-up material and the envelope mass.}
\label{fig3}
\end{figure*}
Due to the partial overlap of the convective zones generated by
the recurrent TPs, the {\it s}-processed material accumulates
within the He-intershell. This occurs for the first convective
shells, in which the production of the {\it s}-process elements
within the $^{13}$C pockets compensate the dilution over the ashes
of the H-shell, which are basically {\it s}-process free (see Fig.
16 of Paper I)\footnote{Actually, this material also presents an
{\it s}-process enrichment, due to previous TDU, but it is
definitely lower than the one characterizing the He-intershell.}.
Afterwards, the dilution effect dominates and in the following TPs
the heavy elements abundances in the He-intershell decrease. Fig.
\ref{fig3} illustrates the ratios between the amount of dredged-up
material per TDU and the residual envelope mass as a function of
the core mass, for the different stellar initial masses. Each
point corresponds to a single TDU episode. Since the variation of
the mass fraction of any isotope in the envelope as a consequence
of a certain TDU episode is proportional to $\Delta M_{\rm
TDU}/M_{\rm ENV}$, such a ratio well represents the efficiency of
the various TDU episodes. It generally increases during the AGB
evolution, with a sharp drop to 0 when the TDU ceases. This ratio
is also generally larger at low $Z$ because the TDU is deeper.
Note that the 3 $M_{\odot}$ model with $Z=0.001$ shows a
substantially smaller $\Delta M_{\rm TDU}/M_{\rm ENV}$ than other
masses with the same metallicity. This model develops a larger
core mass and has a lower $\Delta M_{\rm TDU}$, due to the
shrinking of the He-intershell. Moreover, it has a larger envelope
mass. Its structure is very similar to the one characterizing
Intermediate Mass Stars (IMS) with higher metallicities. The
modelling of IMS is made difficult from the lack of well
constrained (and direct) observational counterparts, even if new
interesting data have been recently presented by \citet{roede}.
Moreover, the computed surface compositions of IMS strongly vary
from author to author, depending on the adopted mass-loss rate,
convection efficiency and opacity set (see, e.g.,
\citealt{vema09}). Unfortunately,
these quantities cannot be fixed one to the time, since they are highly degenerate.\\
\begin{figure*}[tpb]
\centering
\includegraphics[width=\textwidth]{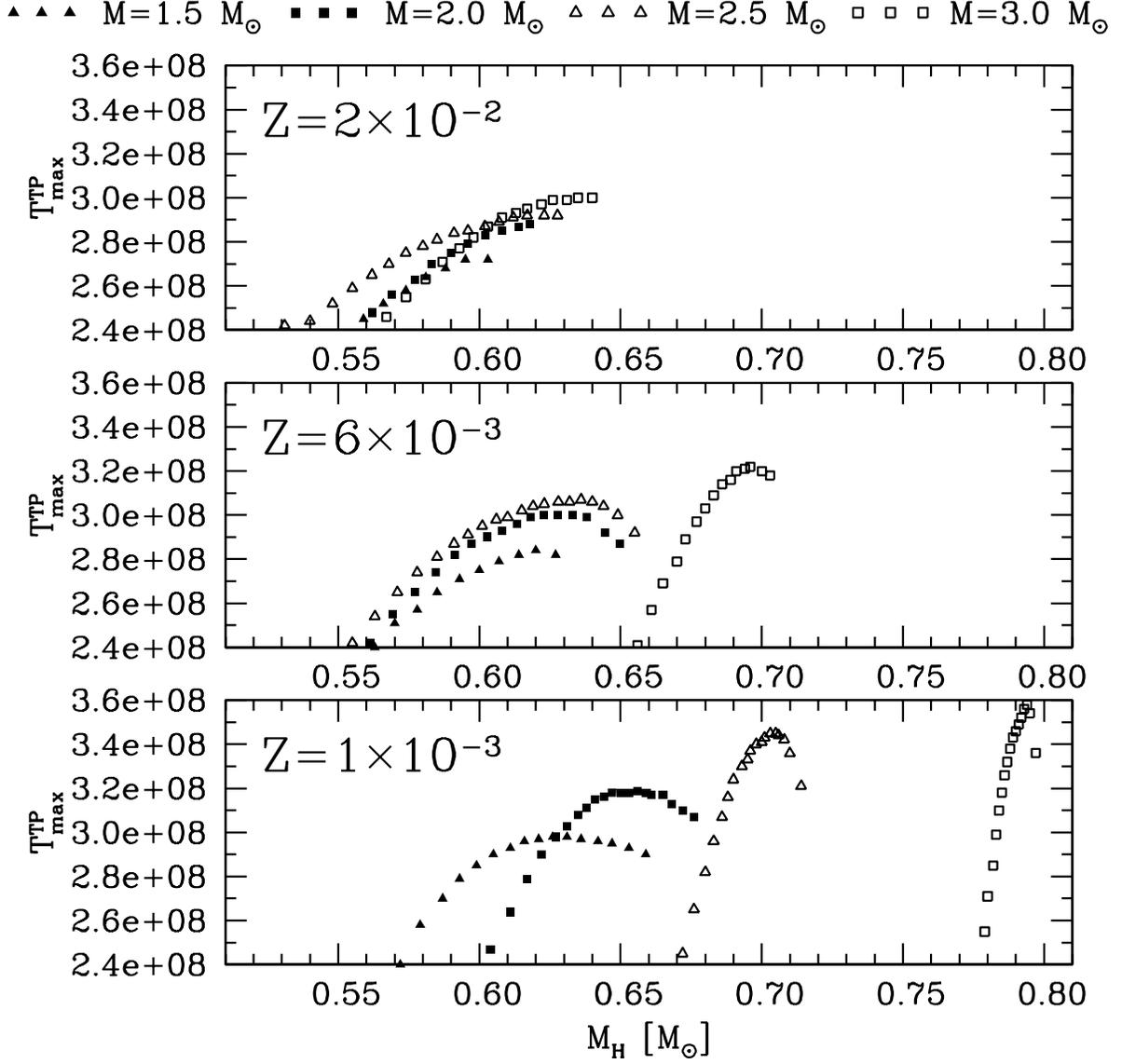}
\caption{As in Fig. \ref{fig2}, but for the maximum temperature
attained at the bottom of the convective zone generated by a TP.}
\label{fig4}
\end{figure*}
As already mentioned in the Introduction, the activation of a
second neutron burst, due to the $^{22}$Ne($\alpha$,n)$^{25}$Mg
reaction, depends on the temperature attained at the base of the
convective zone generated by a TP. This is illustrated in Fig.
\ref{fig4}. For each initial mass a general trend can be seen. The
maximum temperature within the convective zone generated by a TP
progressively grows, reaches a maximum and then slightly decreases
(during the last TPs). As discussed in \citet{stra03a}, this
temperature trend depends on the core mass, the envelope mass and
the metallicity. The maximum temperature is higher for larger core
masses and larger envelope masses, while it is lower for higher
$Z$. As a results, at $Z$=0.02, all stars with $M \le 3 M_\odot$
barely attain the threshold of the $^{22}$Ne($\alpha$,n)$^{25}$Mg
reaction. On the contrary, at $Z$=0.001, in all stars with $M > 2
M_\odot$ the second neutron burst plays a relevant role in the
resulting {\it s}-process nucleosynthesis. As we will show in the
next section, in these models there is a substantial activation of
some {\it s}-process branchings and the corresponding increased
production of branching-dependent isotopes (such as $^{60}$Fe,
$^{86}$Kr, $^{87}$Rb and $^{96}$Zr). In addition, light-{\it s}
(ls) elements\footnote{Those belonging to the first peak of the
{\it s}-process (ls, light elements: Sr-Y-Zr). See next Section.}
are efficiently produced as a consequence of the rapid mixing
taking place in the convective shell. On one hand, this mixing
continuously provides fresh Fe seeds in the innermost (and hotter
layers) of the convective zone. On the other hand, it prevents the
pile up of the synthesized material, which is required to build up
the heaviest species (as it occurs during the radiative $^{13}$C
burning within the pockets).

\section{Nucleosynthesis}\label{nucleo}

Three main peaks are expected in the {\it s}-process distribution.
They correspond to nuclei with a magic number of neutrons (50, 82
and 126), whose binding energy is particularly high and, in turn,
the neutron capture cross-section is particularly low. The average
abundances of {\it s}-process elements around N=50 and N=82
represent respectively the so-called ls (light {\it s}) and hs
(heavy {\it s}) indexes. As in Paper I, we define:
\begin{itemize}
\item{[ls/Fe]=([Sr/Fe]+[Y/Fe]+[Zr/Fe])/3;}
\item{[hs/Fe]=([Ba/Fe]+[La/Fe]+[Nd/Fe]+[Sm/Fe])/4.}
\end{itemize}
The third peak (at N=126) is represented by lead ([Pb/Fe]).
Combining these quantities, the spectroscopic indexes [hs/ls] ,
[Pb/ls] and [Pb/hs] can be derived\footnote{We adopt the usual
spectroscopic notation: [a/b]=$\log$(a/b)-$\log$(a/b)$_\odot$}.
Basing on simple arguments, \citet{cameron} demonstrated that
these spectroscopic indexes basically depend on the number of
neutrons available per iron seed for the first time. In stellar
interiors, however, the physical and chemical conditions may be
more complex. In the case of the radiative $^{13}$C burning taking
place during the interpulse period in low mass AGB stars, the
number of neutrons corresponds to the number of $^{13}$C nuclei in
the pocket. However, most of these neutrons are captured by light
elements (poisons) and do not contribute to the {\it s}-process.
As already specified in Section \ref{agb}, $^{14}$N (the most
abundant poison in our models) partially recycles neutrons.
Nevertheless, as a first order approximation, we may assume that
the number of neutrons per seed available for the {\it s}-process
is given by:
\begin {equation}
 \frac{n(neutrons)-n(poisons)}{n(seeds)}\sim\frac{n(^{13}\rm{C})-n(^{14}\rm{N})}{n(^{56}\rm{Fe})}
\end {equation}\\
The numerator of this equation coincides with the {\it effective}
$^{13}$C introduced in Section \ref{agb}. Within the $^{13}$C
pocket it varies according to the protons profile left by the TDU.
Note that both the $^{13}$C and the $^{14}$N in the pocket share
the same (primary) origin, since both are built from the primary
$^{12}$C\footnote{Primary elements are direct by-products of H or
He burning.}. On the contrary, the iron abundance within the
pocket depends on the initial stellar metallicity.

\begin{figure*}[tpb] \centering
\includegraphics[width=\textwidth]{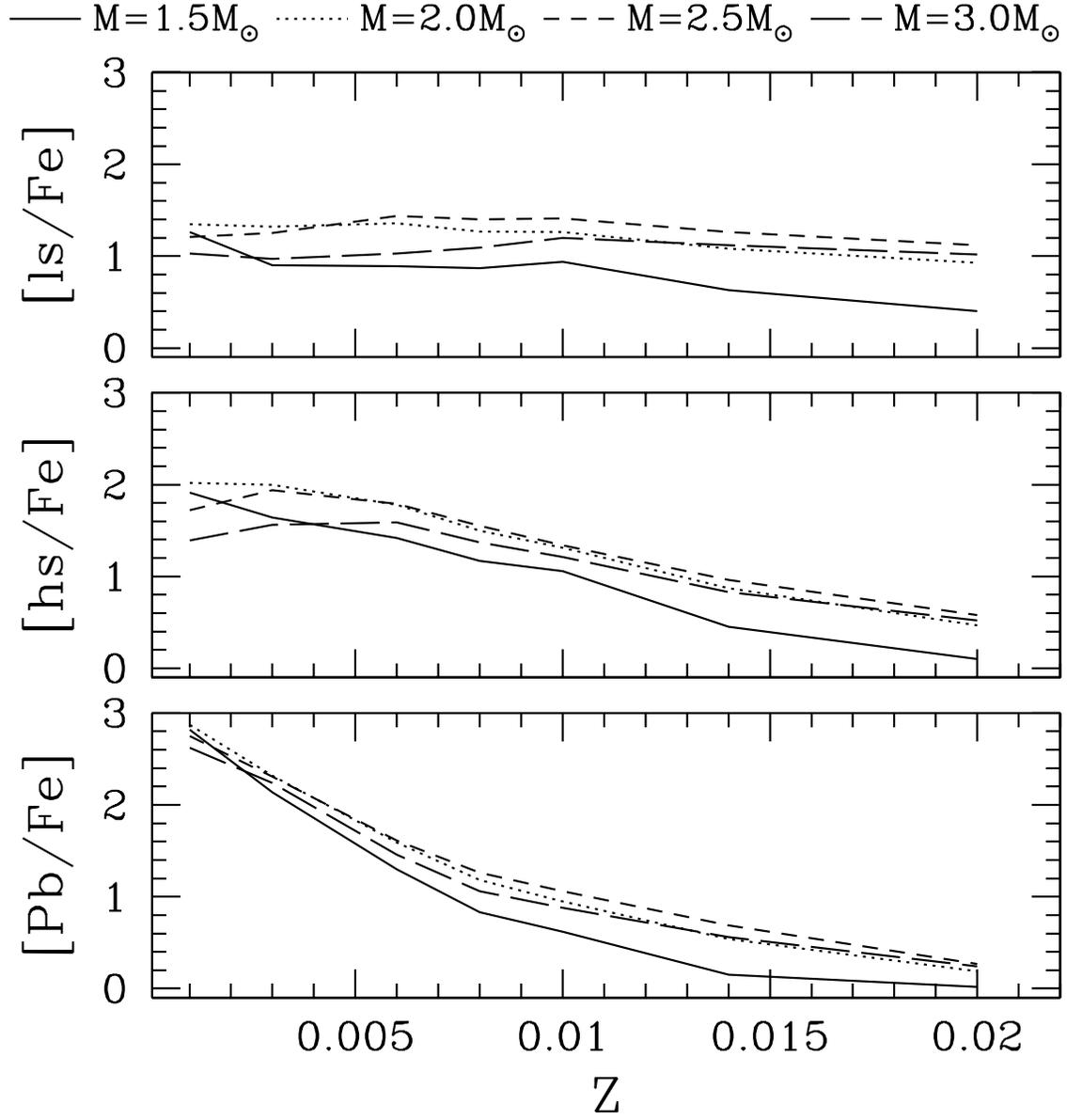}
\caption{Behaviors of the three {\it s}-process peaks ([ls/Fe],
top panel; [hs/Fe], middle panel; [Pb/Fe], lower panel) as a
function of the metallicity  predicted by our models for different
initial masses.} \label{fig5}
\end{figure*}
The behavior of the three {\it s}-process peaks at the last TDU as
a function of the metallicity and for different masses is
illustrated in Fig. \ref{fig5}. For selected masses and
metallicities, the corresponding values are reported in Table
\ref{tab6}. More complete Tables can be downloaded from the FRUITY
web-based platform. We find that [ls/Fe] is barely flat for all
the metallicities, the only 1.5 $M_\odot$ exhibiting a mild
increase when decreasing the metal content. As to [hs/Fe], all
masses show a steep increase with the metallicity. This behavior
is even more evident for [Pb/Fe], which peaks at the lowest
computed metallicity. Given the steep increase of the neutrons per
iron seed with decreasing metallicity (see column 13 in Tables
\ref{tab3}, \ref{tab4} and \ref{tab5}), the heaviest species are
easier synthesized at low $Z$.
\begin{figure*}[tpb] \centering
\includegraphics[width=\textwidth]{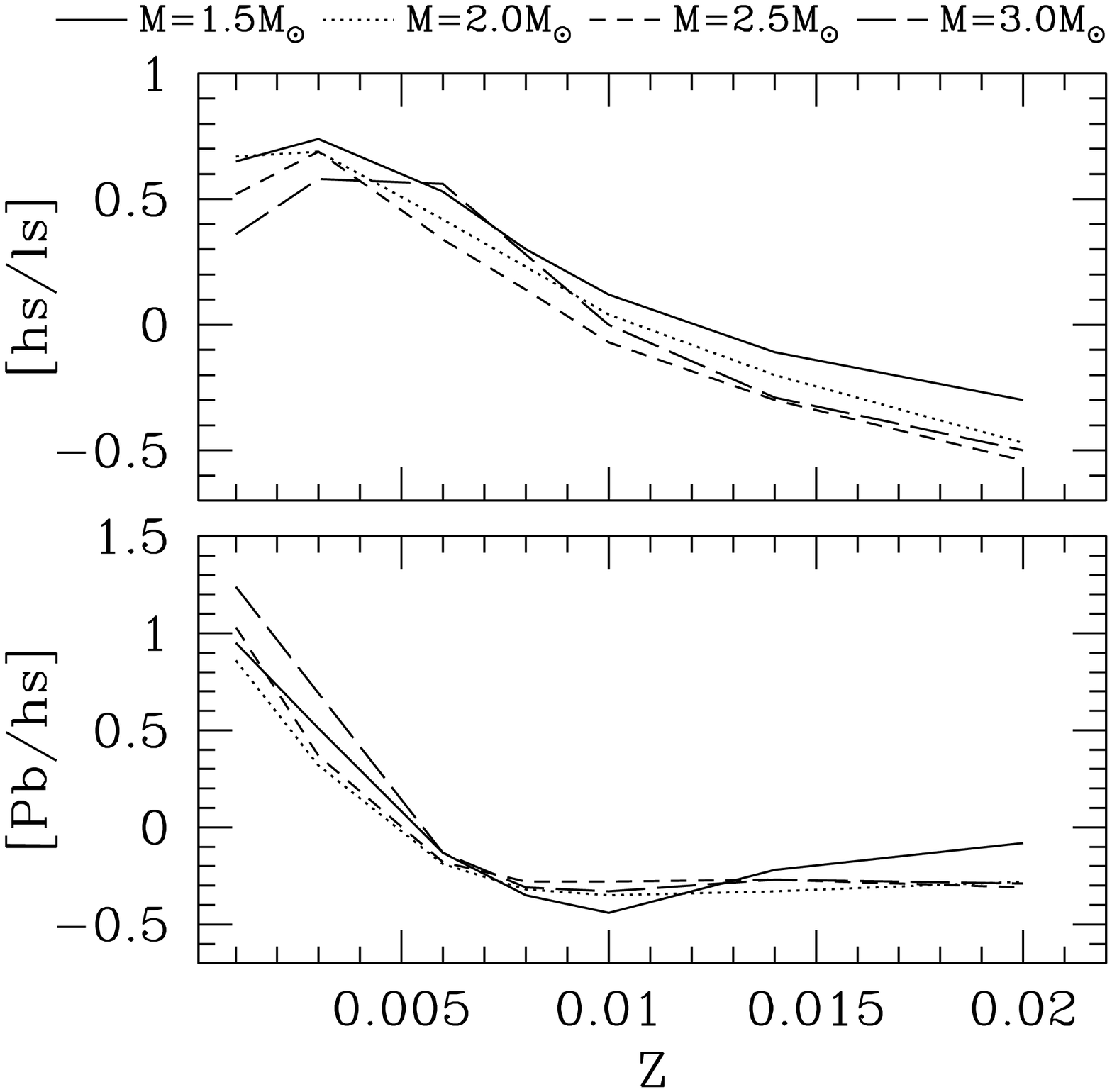}
\caption{As in Fig. \ref{fig5}, but for the {\it s}-process
indexes [hs/ls] (upper panel) and [Pb/hs] (lower panel).}
\label{fig6}
\end{figure*}
In Fig. \ref{fig6} we report the {\it s}-process spectroscopic
indexes [hs/ls] (upper panel) and [Pb/hs] (lower panel). The first
quantity continuously grows with decreasing metallicity, reaching
a maximum at $Z\sim 3\times 10^{-3}$ (eventually decreasing at
lower metallicities), while [Pb/hs] shows a flat profile from
$Z\sim 2\times 10^{-2}$ down to $Z\sim 6\times 10^{-3}$ and then
grows at the lowest metallicities. A full discussion on the
behavior of these quantities as a function of the metal content
can be found in \citet{bi10}. As already outlined, these
quantities mainly depend on the neutrons per iron seed and on the
abundances of neutron poisons. In our models, the amount of
$^{13}$C in the pockets basically depends on the core mass (see
Fig. \ref{fig1}). At fixed metallicity the core masses for
different models do not vary much (apart from the $Z=1\times
10^{-3}$ models, but see below) and, therefore, there is no
particular trend with the initial mass. Our spread at fixed
metallicity reflects the slight variation of the total number of
neutrons per iron seed in each pocket for different initial
masses. This limited spread derives from the fact that, in our
models, only convection is considered as a mixing mechanism: the
introduction of additional mixing phenomena (such as rotation; see
e.g. \citealt{la99} and \citealt{he03}), able to dilute the
available $^{13}$C on a larger mass region, could lead to a lower
neutron to seed ratio (see Section \ref{hsls}).

Other interesting nucleosynthetic features emerge from the
analysis of light elements (see Table \ref{tab6}, in which the
final C/O, C/N and $^{12}$C/$^{13}$C ratios are also given). Among
them, carbon and fluorine show a clear anti-correlation with the
initial metallicity. Carbon is of primary origin, because it is
produced during TPs via the 3$\alpha$ reaction. The production of
fluorine basically depends on the availability of $^{13}$C in the
He-intershell (see Paper I for a detailed description of the
nuclear chain leading to $^{19}$F nucleosynthesis; see also
\citealt{jo92,fo92,mo96,lu04,ab10,ab11}). A non negligible amount
of $^{13}$C is available within the $^{13}$C-pocket and in the
ashes of the H-burning shell. As already discussed in Paper I, the
former is weakly dependent on metallicity. The latter directly
scales with the CNO abundances in the envelope, which, in turn,
depend on the (primary) $^{12}$C dredged-up during all previous
TDU episodes.
The resulting fluorine can be roughly considered a primary element. \\
The nitrogen surface enrichment is mainly a consequence of the FDU
and, at the lowest metallicities, of the H-shell erosion by the
envelope during TDU episodes. This latter contribution is not
appreciable at higher metallicities
because of the already large $^{14}$N surface abundance. \\
The neon surface enrichment derives by the dredge-up of $^{22}$Ne,
which is synthesized during TPs via the nuclear chain
$^{14}$N($\alpha$,$\gamma$)$^{18}$F($\beta^+\nu$)$^{18}$O($\alpha$,$\gamma$)$^{22}$Ne.
Direct evidence of neon production has recently been detected in
Galactic planetary nebulae \citep{mi10}. Since the nitrogen
abundance in the He-intershell depends on the quantity of carbon
previously dredged-up (due to the conversion of $^{12}$C to
$^{14}$N via the CNO cycle), $^{22}$Ne is a primary
element (thus its surface overabundance is expected to increase at low metallicities). \\
In our models, sodium can be synthesized via proton or neutron
captures (see also \citealt{mo99}). A sodium pocket forms via the
$^{22}$Ne(p,$\gamma$)$^{23}$Na reaction in the upper part of the
He-intershell, where protons are left by the receding convective
envelope after TDU episodes (see Paper I). Moreover, $^{23}$Na is
synthesized via the
$^{22}$Ne(n,$\gamma$)$^{23}$Ne($\beta^-\nu$)$^{23}$Na nuclear
chain, during both the radiative burning of the $^{13}$C-pocket
and
the convective $^{22}$Ne burning in the TPs. \\
Magnesium isotopes follow different nucleosynthetic paths.
$^{24}$Mg, which is the most abundant magnesium isotope in the
Sun, is marginally produced via neutron captures on $^{23}$Na
during the radiative burning of the $^{13}$C pocket and it is
weakly destroyed during TPs. On the contrary, $^{25}$Mg and
$^{26}$Mg are mainly created during TPs via neutron captures
processes and via the $^{22}$Ne($\alpha$,n)$^{25}$Mg and the
$^{22}$Ne($\alpha$,$\gamma$)$^{26}$Mg reactions (which contribute
to the $^{25}$Mg and $^{26}$Mg nucleosynthesis, respectively). In
our models we generally find a low final surface magnesium
enrichment, with an increasing trend toward low metallicities.
This derives from the larger relative contribution of $^{25}$Mg
and $^{26}$Mg, due to the increased efficiency in the activation
of the $^{22}$Ne($\alpha$,n)$^{25}$Mg and
$^{22}$Ne($\alpha$,$\gamma$)$^{26}$Mg reactions (see Section
\ref{agb}).

All our models become C-rich ({\it i.e.} their final C/O$>$1), but
the 1.5 $M_\odot$ model at $Z=2\times 10^{-2}$. As already
mentioned, non convective mixing episodes (the so-called
extra-mixing) for both the RGB and the AGB phases are not
considered in these models. The origin of this extra-mixing is far
from being established on theoretical grounds. The precise
estimate of its efficiency during both the giant branches can be
obtained only by comparing parameterized models with observational
constraints \citep{cz07,eg08,no08,dpm09,bu10,de10,palmerini,de11}.
The extra-mixing is expected to affect the CNO abundances, with a
substantial reduction of the $^{12}$C/$^{13}$C isotopic ratio (see
also \citealt{leb08,lederer}) and, eventually, a lowering of the
C/O and C/N ratios. For this reason, special attention has to be
paid when comparing the surface abundances of these elements (or
isotopes) with their observed counterparts.

A selection of net stellar yields for different masses and
metallicities is reported in Tables \ref{tab7}, \ref{tab8} and
\ref{tab9} (full Tables are available on line at FRUITY web
pages). These data represent the first homogeneous set of light
and heavy elements theoretical yields for AGB stars. The
completeness of our database could help to reduce the
uncertainties affecting Galactic chemical evolutionary models, as
underlined by \citet{romano}. At high metallicities ($Z\sim
2\times 10^{-2}$), the largest absolute yields (for both light and
heavy elements) come from the 2.5 $M_\odot$ and the 3.0 $M_\odot$.
At lower metallicities ($Z\sim 6\times 10^{-3}$) the principal
contribution derives, for all chemical species, from the 2.5
$M_\odot$ model. At $Z\sim 10^{-3}$, the initial mass of the
principal polluter further decreases, presenting the largest
yields in the 2.0 $M_\odot$ mass. Weighing the yields with an
Initial Mass Function (IMF), for example the one proposed by
\citet{salpe}, the contribution from the 2.0 $M_\odot$ stars
dominates at all metallicities. \\
The amount of processed material from various masses at different
metallicities results from a combination of many factors, as the
core mass, the efficiency of TDU, the envelope mass, the number of
TPs, the thickness of $^{13}$C pockets and the surface temperature
(which drives the mass-loss rate). As a rule of thumb, a
progressive decrease of the metallicity leads to a reduction of
the yields from stars with $M\sim (2.5 \div 3.0) M_\odot$, making
stars with $M\sim (1.5 \div 2.0) M_\odot$ the principal polluters
of the ISM. Note that, as already pointed out, the explicit
treatment of physical phenomena currently not accounted for (such
as rotation) could significantly modify these quantities and, in
particular, the surface distributions of the heavy elements.

In order to better interpret the nucleosynthetic paths described
above, we highlight some general rules characterizing our models:
\begin{itemize}
\item{the main neutron source is represented by the radiative
burning of $^{13}$C within the pockets;} \item{the nucleosynthesis
occurring in the first 4$-$5 pockets mainly determines the final
surface distributions;} \item{the contribution of the additional
$^{22}$Ne($\alpha$,n)$^{25}$Mg neutron burst is generally
negligible;} \item{the efficiency of the TDU increases with
decreasing the metallicity;} \item{at fixed metallicity, the
surface chemical enrichment basically depends on the total amount
of dredged-up material and on the mass extension of the convective
envelope\footnote{See Section \ref{uncer} for a discussion about
the effects induced by different mass-loss rates on the final
elemental surface abundances.}.}
\end{itemize}
However, there are a few exceptions. As already described in Paper
I, the first $^{13}$C-pocket of the 2 $M_\odot$ models does not
completely burn during the radiative interpulse period and some
$^{13}$C is ingested and burnt convectively in the following TP.
We confirm this finding for the other masses. This episode has no
effects on the physical evolution of the TP, while it has
interesting effects from an isotopic point of view. In particular,
the large neutron densities attained (see Paper I) allow the
nucleosynthesis of some neutron-rich isotopes, such as the short
lived $^{60}$Fe. \\
A characteristic feature of our 3 $M_\odot$ model at $Z=10^{-3}$
is represented by the occurrence of the so-called Hot-TDU
\citep{gosi,he04,cl08,lau09}. In fact, for this model proton
captures occur at the base of the convective envelope during the
TDU phase. In Paper I, we already found the Hot-TDU to occur in
the M=2 $M_\odot$ model with $Z=10^{-4}$. As a first consequence,
the model shows a definitely lower $^{12}$C/$^{13}$C isotopic
ratio with respect to other masses (see Table \ref{tab6}). Note
that the occurrence of the Hot-TDU depends on the mixing algorithm
we applied at the inner border of the convective envelope (see
Section \ref{code}). The calibration of its $\beta$ parameter has
been done in order to maximize the {\it effective} $^{13}$C in the
pockets and, consequently, the resulting {\it s}-process
nucleosynthesis. Actually, it is not clear if the same calibration
can be also applied to AGB stars developing a larger core mass, as
our 3 $M_\odot$ model at $Z=10^{-3}$. With respect to the other
calculated masses, this model also reaches higher temperatures
during TPs, which imply a more efficient activation of the
$^{22}$Ne($\alpha$,n)$^{25}$Mg neutron source. This leads to a non
negligible contribution to [ls/Fe]. Consequently, the [hs/ls]
index decreases, leading to a larger spread of this
quantity when compared to the other metallicities. \\
It is worth to mention that the nucleosynthesis of some isotopes
(the ones whose production requires a high neutron density) is
favored by the activation of the $^{22}$Ne neutron source. In
Table \ref{tab10} we list some isotopic ratios involved in
branchings along the {\it s}-process path. The dependence on the
metallicity has already been discussed in Paper I, demonstrating
that these ratios increase with decreasing metallicity, due to the
increased contribution to the {\it s}-process nucleosynthesis of
the $^{22}$Ne($\alpha$,n)$^{25}$Mg. The dependence of the
corresponding isotopic ratios on the initial mass is clearly shown
in Table \ref{tab10}. With the exception of the most metal-rich
models, the 3 $M_\odot$ models show the largest ratios, confirming
again that the $^{22}$Ne($\alpha$,n)$^{25}$Mg neutron source is
particularly important for the more massive AGB stars.

\section{Theoretical uncertainties}\label{uncer}

When dealing with observations, it is mandatory to perform a
detailed discussion of the errors introduced by the instruments
and the techniques adopted to analyze data. A similar procedure
should be followed when computing theoretical stellar models. In
this case, error-bars cannot be derived (due to the high degree of
degeneracy among the input parameters), but the sensitivity of the
model with respect to changes in the input physical parameters can
be evaluated. A similar analysis was already performed in Paper I
for a 2~$M_\odot$ AGB model with $Z=10^{-4}$. Considering the
larger metallicities of the models presented in this work, we
focus on a $M = 2 M_\odot$ with $Z=6\times 10^{-3}$ (hereafter
FRANEC Standard Model, FSM). We assume as our FSM the model
computed with the following set of parameters:
\begin{itemize}
\item{the free parameter of the mixing length theory is set to
$\alpha_{\rm FSM}$=2.15 (see Paper I);}  \item{the AGB mass-loss
rate ($\dot M_{\rm FSM}^{\rm AGB}$) is empirically calibrated on
the observed period-luminosity relation (see \citealt{stra06} for
details);} \item{the free parameter that determines the decreasing
profile of the convective velocities at the base of the envelope
is set to $\beta_{\rm FSM}$=0.1 (see Paper I).}
\end{itemize}
In order to estimate the sensitivity of our models to the
afore-listed inputs we calculate six test models. We vary the
mixing length parameter $\alpha$, the mass loss law and the $\beta
$ parameter which governs the efficiency of the velocity profile
at the border of the convective envelope:
\begin{enumerate}
\item{$\alpha_1=\alpha_{\rm FSM}+0.35=2.5$;}
\item{$\alpha_2=\alpha_{\rm FSM}-0.35=1.8$;} \item{$\dot M_1^{\rm
AGB}=\dot M_{\rm FSM}^{\rm AGB}*2$;} \item{$\dot M_2^{\rm
AGB}=\dot M_{\rm FSM}^{\rm AGB}*0.5$;}\item{$\beta_1=\beta_{\rm
FSM}+0.02=0.12$;} \item{$\beta_2=\beta_{\rm FSM}-0.04=0.06$.}
\end{enumerate}
A variation of the mixing length parameter $\alpha$ affects not
only the temperature run in the more external layers of the star,
where the local temperature gradients largely deviate from the
adiabatic value, but also the velocity profile of the convective
bubble in the whole convective envelope (see \citealt{pi07}). \\
We evaluate the effects of a variation of the AGB mass-loss rate
by multiplying or dividing its efficiency by a factor 2. At fixed
pulsational period, the resulting theoretical mass loss laws
reproduce the minimum and the maximum mass-loss rates observed in
Galactic AGB stars (see Figure 3 of
\citealt{stra06}).\\
Finally, we present two models with a different $\beta$ parameter
in order to estimate the effects induced by a change in the slope
of the exponentially decreasing profile of convective velocities
at the base of the envelope. We choose values leading to a
reduction of the {\it effective} $^{13}$C within the pockets by
roughly a factor 4 with respect to our FSM (see Figure 7 of Paper I).\\
In Table \ref{tab11} we report, for our FSM, the final core mass
($M_{\rm H}$), the mean bolometric magnitude of the AGB C-rich
phase ($\overline{\rm M}$$_{\rm bol}^{C}$)\footnote{Calculated as
the average luminosity weighted over the time spent by the star on
the AGB when C/O$>$1.}, the duration of the AGB C-rich phase
($\tau_C$), the $^{12}{\rm C}$ yield (as representative of
primary-like light isotopes), the $^{89}{\rm Y}$, $^{139}{\rm La}$
and $^{208}{\rm Pb}$ yields (as representative of the three {\it
s}-process peaks) and the final [hs/ls] ratio. For the test cases,
we list normalized
percentage differences with respect to the FSM.\\
The physical quantities listed in Table \ref{tab11} (final core
mass, mean bolometric magnitude of the C-rich phase and duration
of the C-rich phase) mildly depend on the choice of the input
parameters. Maximum variations are below 30\%. This makes our
theoretical models quite robust when comparing with the observed
counterparts (see Section \ref{cstarlum}). \\
The test with an enhanced $\alpha$ parameter and the test with a
reduced mass loss rate present similar trends: the mean bolometric
magnitude of the C-rich phase decreases (and, thus, the mean
luminosity increases), while the C-rich phase duration and the
chemical yields increase. The reason is twofold. In fact, a higher
$\alpha$ parameter implies more efficient TDUs and a slightly
higher surface temperature, which translates into a milder mass
loss rate. An opposite trend is found for the test with a reduced
$\alpha$ parameter which, thus, resembles the test with an
increased mass loss rate. Note that, for the aforementioned test
models, the heavy elements yields roughly scale as the light ones
and the final chemical patterns show minor
changes with respect to the FSM (see their [hs/ls] ratios). \\
A different prescription for the mass loss law leads to negligible
differences in the {\it s}-process abundances because the first
$^{13}{\rm C}$ pockets largely determine the final heavy elements
enrichment in the envelope. For the same reason, minor changes in
the relative abundances of {\it s}-process elements are found in
the $\alpha$-varied tests, even if a variation of this parameter
leads to different TDU efficiencies. However, a further aspect has
to be analyzed. Within the framework of the mixing length theory,
the value of the convective velocities directly scales with
$\alpha$. Thus, a variation of such a parameter should also modify
the efficiency of the mixing algorithm we adopt at the base of the
convective envelope. However, we verified that this has minor
effects on the formation of the $^{13}$C pockets, with variations
lower than 15\% in the {\it effective} $^{13}{\rm C}$. This is not
the case for the two tests with a different $\beta$ parameter. In
fact, these models present complete different heavy elements
distributions with respect to the FSM. In both cases there is a
sensible decrease of the absolute surface abundances (as proved by
the yields) and a change in the final shape of the {\it s}-process
distribution (as showed by the [hs/ls] ratios). This is due to the
fact that the amount of neutrons available for the nucleosynthesis
of heavy species within the $^{13}{\rm C}$ pockets decreases in
both cases. The reasons are, however, different. In the
$\beta$=0.06 case, the total amount of $^{13}$C within the pockets
strongly decreases due to the steeper decreasing profile of
convective velocities. Instead, in the $\beta$=0.12 case the
$^{13}$C pocket is completely overlapped with the $^{14}$N pocket.
This acts as a neutron poison via the resonant
$^{14}$N(n,p)$^{14}$C reaction, thus penalizing the heavy elements
nucleosynthesis (see Section \ref{nucleo}).

\section{Comparison with other stellar models}\label{modteo}

In previous Sections we discussed in detail our theoretical
models, by analyzing their physical and chemical properties as a
function of the initial mass and metal content. Moreover, we
evaluated the uncertainties deriving from different choices of the
input parameters.
\begin{figure*}[tpb] \centering
\includegraphics[width=\textwidth]{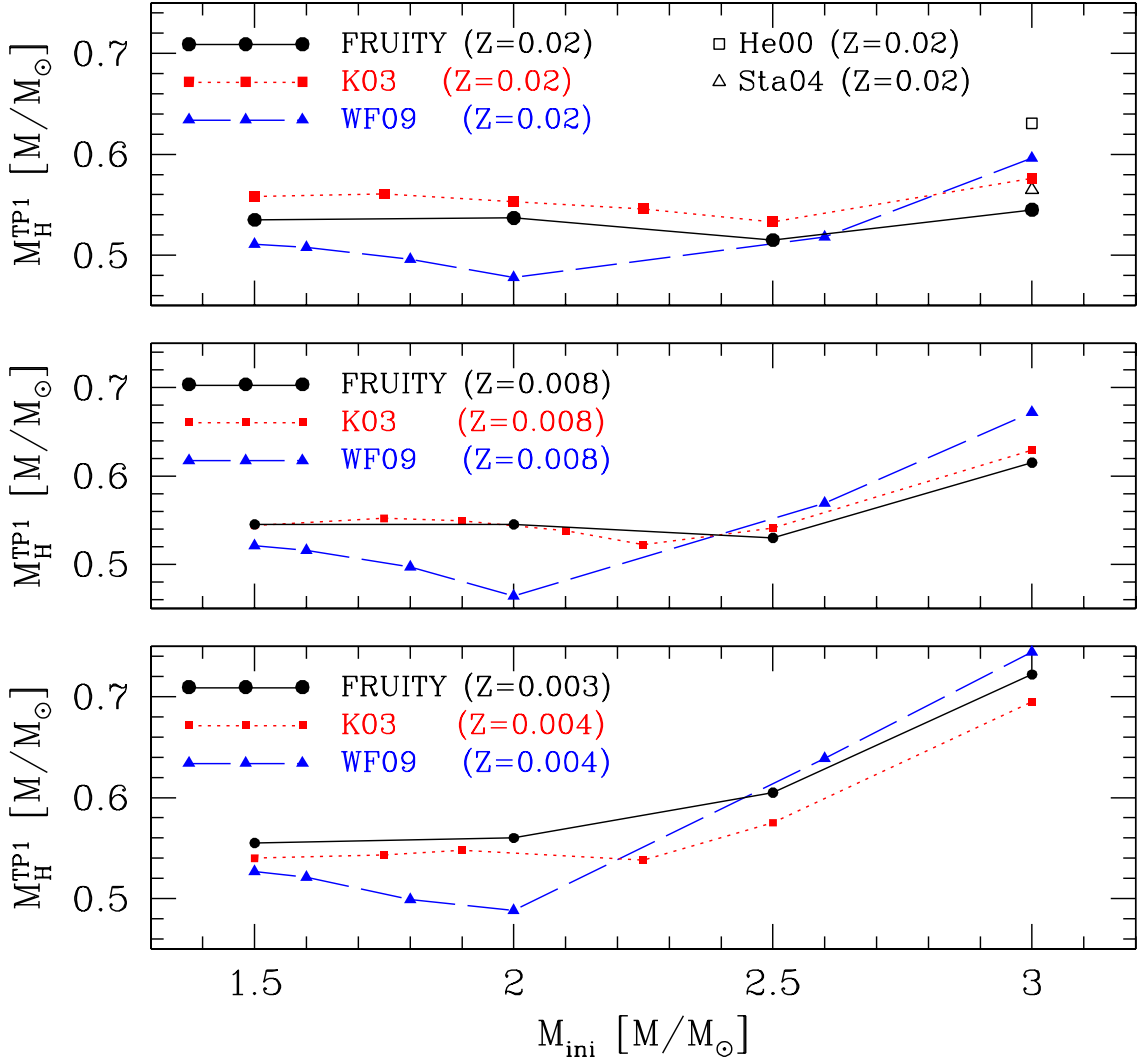}
\caption{Core mass at the first TP as a function of the initial
stellar mass at various metallicities and for different authors
(FRUITY: this paper; K03: \citealt{kakka03}; WF09:
\citealt{weiss}; He00: \citealt{he00}; Sta04: \citealt{sta04}).}
\label{fig6bis}
\end{figure*}
In this Section we present a comparison between our results and
other evolutionary stellar models available in the literature. A
similar comparison is in general not straightforward, since the
various codes significantly differ in many of the adopted input
physics (treatment of convection, definition of convective
borders, mass-loss law, initial chemical abundances, opacities,
equation of state and nuclear network). As already discussed in
Section \ref{code}, during the pre-AGB phase our models are very
similar to those presented in \citet{dom99}. We refer to that
paper for a comparison between different codes. As discussed in
\citet{stra03a}, a key physical quantity to understand AGB
evolution is the core mass at the beginning of the TP-AGB phase
(we identify it as the core mass at the first TP in which the
helium luminosity exceeds 10$^4 L_\odot$, $M^{\rm TP1}_{\rm H}$).
In Fig. \ref{fig6bis} we plot our $M^{\rm TP1}_{\rm H}$ for three
selected metallicities ($Z=0.02$, $Z=0.008$ and $Z=0.003$) and
compare them to models from \citet{kakka03} (hereafter K03) and
\citet{weiss} (hereafter WF09). For the $M=3 M_\odot$ model with
$Z=0.02$ we also report data from \citet{he00}  (hereafter
He00) and \citet{sta04} (hereafter Sta04). \\
The value of $M^{\rm TP1}_{\rm H}$ basically depends on the core
mass attained at helium ignition ($M_{\rm He}^1$ in Table
\ref{tab2}) and on the duration of core He-burning. $M_{\rm He}^1$
is built during the RGB phase for an initial mass lower than 2
$M_\odot$. $M_{\rm He}^1$ depends on several input physics, like
the neutrino cooling rate, the $^{14}$N(p,$\gamma$)$^{15}$O
reaction rate and the equation of state for the degenerate matter
of the RGB core (including the Coulomb correction). For higher
initial masses, which develop a sizeable convective core during
central H-burning, the treatment of convection (overshoot or not)
may also affect $M_{\rm He}^1$. On the other hand, the duration of
the core He-burning phase depends on the treatment of convection
(semi-convection or mechanical overshoot) and on the adopted
values for the key reaction rates 3$\alpha$ and
$^{12}$C($\alpha$,$\gamma$)$^{16}$O. In addition, we stress that
the difference in the initial H and He abundance has a
significative influence on the whole stellar model evolution.\\
Our models are in fair agreement with those presented by K03, who
operated similar choices (no core overshoot during core H-burning
and semi-convection during core He-burning). A similar $M^{\rm
TP1}_{\rm H}$ is also attained by Sta04, who also did not apply
convective overshoot. Note that Sta04 simultaneously solve the
equations of the chemical evolution (including the convective
mixing) and those of the physical evolution. Such numerical
algorithm could mimic a treatment of semi-convection similar to
the one we adopt. The reason of the notable difference between our
models and the WF09 ones for the $M=2 M_\odot$ masses is not
clear. We note that, for these masses, the duration of their core
He-burning phase is larger than ours. Such an occurrence suggests
that their $M_{\rm He}^1$ should be lower than ours (lower He core
masses imply lower luminosities and then larger He-burning
lifetimes). Unfortunately, WF09 did not report $M_{\rm He}^1$. On
the contrary, their $M=3 M_\odot$ models have larger $M^{\rm
TP1}_{\rm H}$ than us and K03. This could derive from the
application of convective overshoot during central H-burning. Note
that an even larger $M^{\rm TP1}_{\rm H}$ is attained by He00, who
also applied overshoot during both core
hydrogen and helium burnings.\\
In order to compare the TDU efficiency among different codes, in
Fig. \ref{fig6ter} we report the variation of the $\lambda$ factor
as a function of the core mass for our AGB model with $M=3
M_\odot$ and $Z$=0.02. We plot the same quantity for a model
computed with an older version of the FRANEC code \citep{stra97}
(hereafter Stra97) and for calculations by other authors (He00;
Sta04; \citealt{kakka07}, hereafter K07).
\begin{figure*}[tpb] \centering
\includegraphics[width=\textwidth]{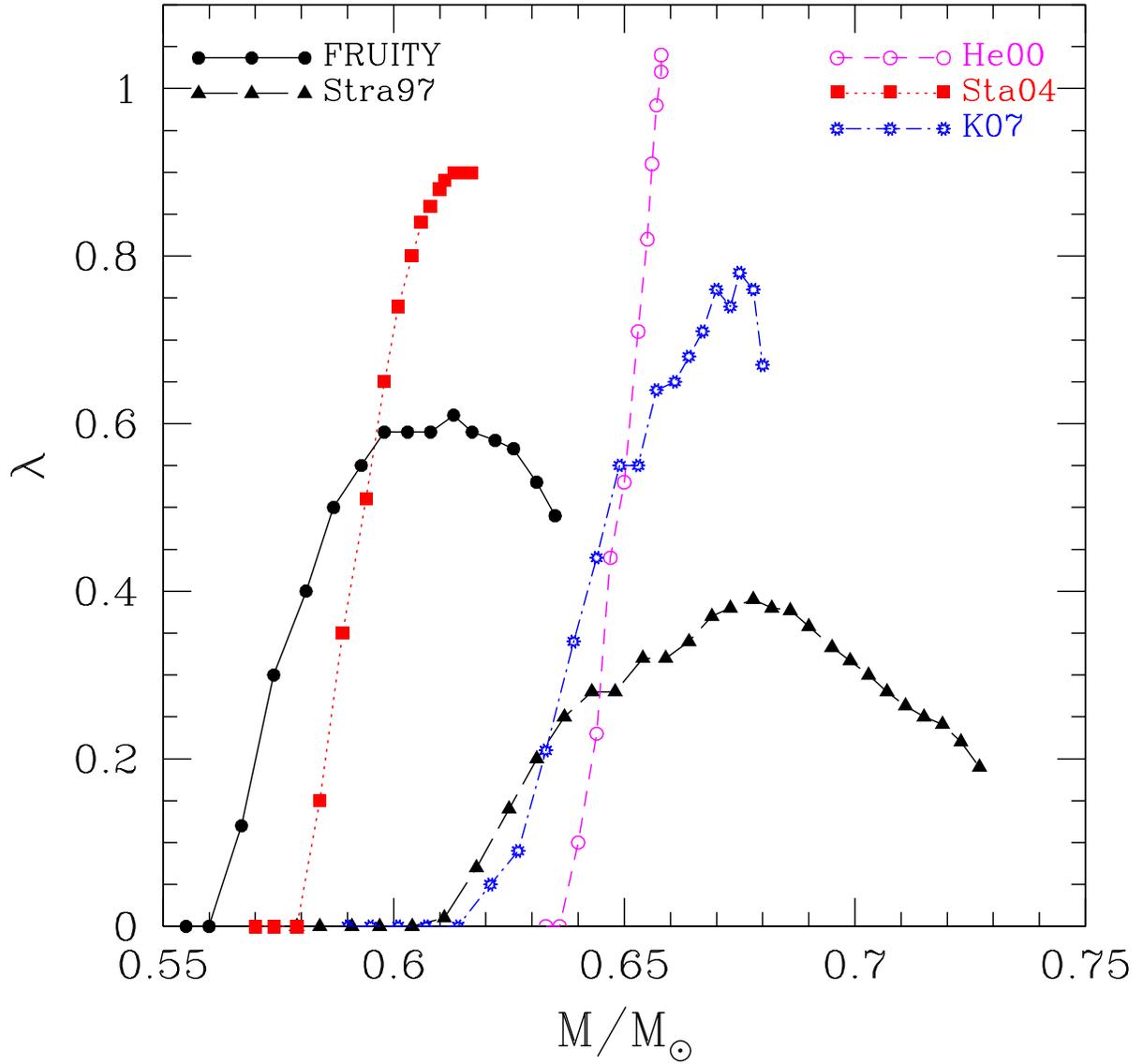}
\caption{$\lambda$ factor as a function of the core mass for an
AGB model with $M=3 M_\odot$ and $Z$=0.02 as reported by different
authors (FRUITY: this paper; Stra97: \citealt{stra97}; He00:
\citealt{he00}; Sta04: \citealt{sta04}; K07: \citealt{kakka07}).}
\label{fig6ter}
\end{figure*}
The TP-AGB phase of the Stra97 model was computed without the
exponentially decaying profile of convective velocities at the
base of the convective envelope (see Section \ref{code}). As a
by-product, the TDU efficiency increases (in fact the maximum
$\lambda$ grows from their 0.4 to the current 0.6) and, in
addition, the TDU occurs earlier (see also the K07 model). Further
differences can be ascribed to the adopted mass-loss law. In fact,
as shown by \citet{stra03a}, the TDU efficiency, and then
$\lambda$, substantially depends on the residual envelope mass.
The K07 model is calculated with a milder mass-loss rate than the
one adopted by Stra97 (Reimers' formula with $\eta$=3 during the
AGB phase). Thus, for the same core mass, their model has a larger
envelope mass and, in turn, larger $\lambda$ values. For the same
reason, the Sta04 model, which is calculated without mass-loss,
attains larger $\lambda$ values than ours. Note that the most
efficient TDU is obtained by He00, even if he included a moderate
mass-loss (Reimers' formula with $\eta$=1 during the AGB phase).
However, in his model the convective overshoot is also applied at
the base of the convective shells, this fact causing stronger TPs
and, consequently, more efficient TDUs (see \citealt{lu03}).
Finally, further differences could be ascribed to the choice of
the $^{14}$N(p,$\gamma$)$^{15}$O
reaction rate \citep{stra00,heau}.\\
The \citet{foca} and the \citet{kakka10} (hereafter K10) are the
only papers we found in the literature in which yields from full
stellar evolutionary AGB models have been discussed (we exclude
synthetic yields from our analysis; see e.g. \citealt{groe} and
\citealt{mari}). We compare the yields of our $M=3 M_\odot$ and
$Z$=0.02 model with the one presented by K10, by considering her
model including a partial mixing zone of $2\times 10^{-3} M_\odot$
after any TDU episode. For its physical characteristics, it is the
most similar model to ours; as a matter of fact, the cumulatively
dredged-up masses are nearly the same ($M_{\rm TDU}^{\rm
K10}=7.93\times 10^{-2} M_\odot$; $M_{\rm TDU}^{\rm
FRUITY}=7.54\times 10^{-2} M_\odot$, see Table \ref{tab3}).
Note that while we have computed a full network stellar
model calculations, the yields presented by K10 are obtained by
means of a post-process calculation. In Table \ref{tab12} we
report a selection of net stellar yields. The analysis is limited
to light isotopes since heavy elements are not included in K10
calculations. For each isotope we report our net yield (labelled
as FRUITY), the corresponding value tabulated in K10 and the
percentage difference. In spite of the many physical differences,
the resulting yields show a reasonable agreement. The differences
concerning $^{20}$Ne, $^{24}$Mg and $^{28}$Si basically derive
from the small AGB net production (or destruction) as compared to
their initial abundances.

\section{Comparison with observations}\label{obs}

In the previous Section, we presented a comparison among our
models and other models available in the literature. As we have
shown, the different results can be ascribed to the different
adopted input physics. This comparison, however, does not indicate
the most reliable input physics or numerical scheme to be
preferred. In order to do that, we have to be guided by
observations. This is the purpose of the present Section, where we
compare our theoretical models to observational counterparts. Here
we concentrate on the Luminosity Function of Carbon Stars and the
{\it s}-process spectroscopic indexes at different metallicities.

\subsection{Luminosity Function of Carbon Stars}\label{cstarlum}

As already outlined in previous Sections, the majority of our
TP-AGB models present final C/O ratios larger than 1; thus, their
observational counterparts are C-rich stars. The Luminosity
Function of Carbon Stars (hereafter LFCS) represents a good test
indicator for our theoretical prescriptions, given that it links a
physical quantity (the luminosity) with the chemistry of the model
(its surface carbon abundance). By comparing our theoretical LFCS
with observational data, we can simultaneously check if the final
structure (basically the core mass) and the surface chemical
abundances of our models (which depend on the TDU efficiency
and the mass loss law) are reliable or not. \\
\citet{ib81} firstly highlighted the incapability of theoretical
AGB models to reproduce the LFCS of Magellanic Clouds. In his
models, TDU was found to occur for core masses larger than 0.6
$M_\odot$ (thus corresponding to AGB stars of Intermediate Mass: 4
$<M/M_\odot<$ 7). However, in these objects the temperature at the
base of the convective envelope is high enough to activate an
efficient hydrogen burning. This phenomenon, known as Hot Bottom
Burning \citep{su71,ib73,kakka03,veda}, basically prevents these
stars from becoming C-rich, due to the conversion of the
dredged-up carbon into nitrogen\footnote{Note that none of the
models presented in this paper suffers for Hot Bottom Burning.}.
Therefore, a discrepancy between theory and observations arose,
called "carbon stars mystery" \citep{ib81}. Later, \citet{la89}
and \citet{stra97} found TDU to occur in stars with masses as low
as 1.5 $M_\odot$, thus allowing the possibility to form C stars at
lower luminosities. Notwithstanding, \citet{izzard} concluded that
full stellar evolutionary models cannot reproduce the Magellanic
Clouds LFCS, being the evolutionary models too luminous with
respect to the observed distributions. These authors concluded
that TDU must operate earlier on the AGB or at lower core masses.
More recently, two groups \citep{stan05,magi} stated to have
reproduced the LFCS of Magellanic Clouds. \citet{stan05} achieved
a good agreement for the Large Magellanic Cloud and suggested a
shift toward lower masses for stars experiencing TDU in order to
reproduce the LFCS of the Small Magellanic Cloud. A full match for
both the Magellanic Clouds LFCS has been found by \citet{magi}.
However, these authors used synthetic stellar models and their
agreement derives from the ad-hoc calibration of stellar
properties during the TP-AGB phase (in
particular the onset and efficiency of the TDU process). \\
A different approach has been followed by \citet{guanda}, who
performed an analysis of Galactic carbon stars at infrared (IR)
wavelengths. They suggested that the inconsistency between theory
and observations is due to the underestimation of the IR
contribution of these objects. In fact, since these stars are very
cool, they emit a large part of their radiation at IR wavelengths.
Actually, these authors found a distribution peaked at M$_{\rm
bol}=-5$, thus in better agreement with model predictions, with a
non negligible luminous tail up to M$_{\rm bol}=-6$. It has to be
stressed that their sample contains a large fraction of stars
(about 45\%, see their Table 3) whose distance is poorly known and
should be better determined. The use of recently improved
observational data (in particular distances) could lead to
significant changes in
their observed LFCS.\\
\begin{figure*}[tpb] \centering
\includegraphics[width=\textwidth]{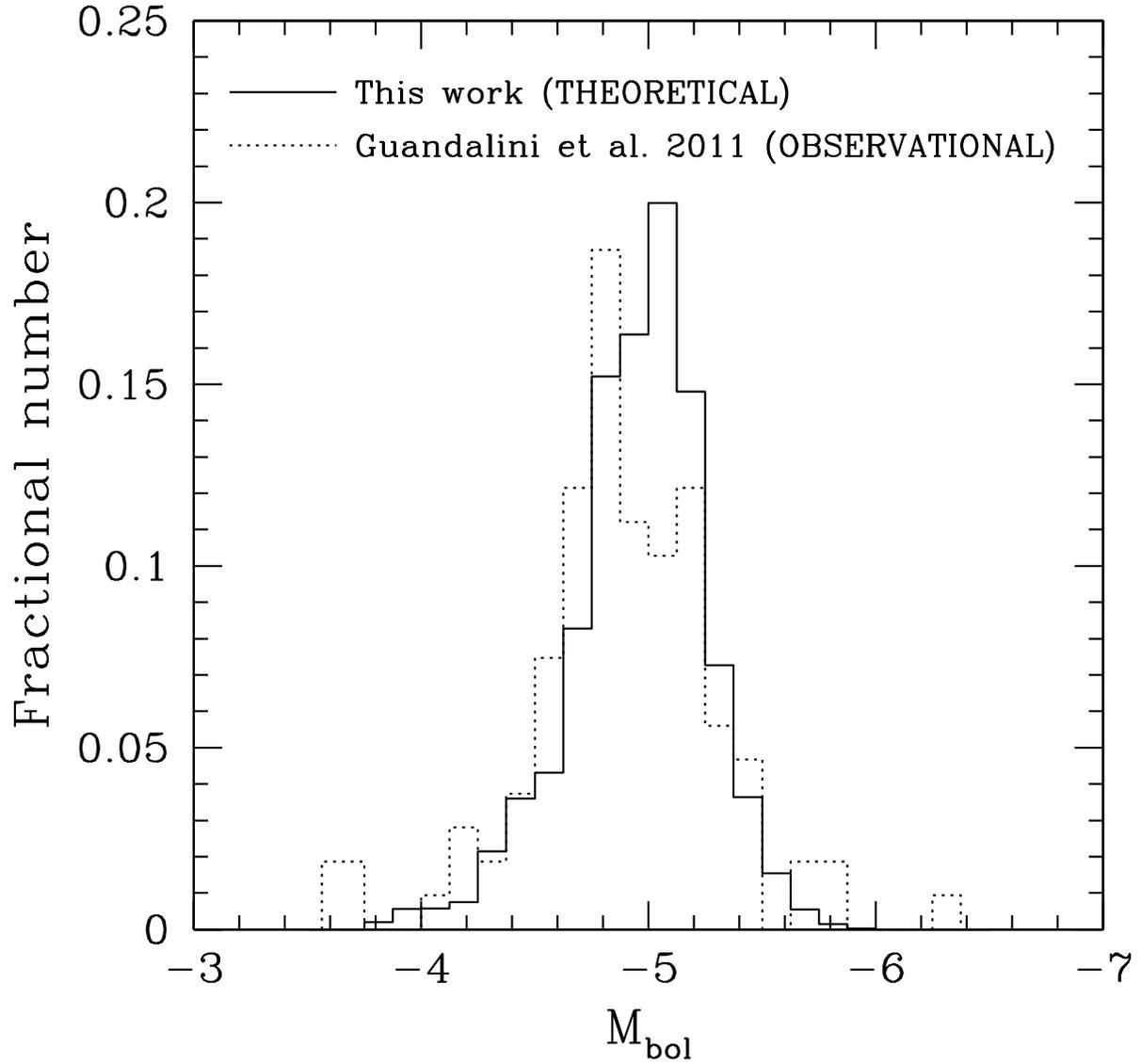}
\caption{Theoretical (solid histogram) Luminosity Function of
Galactic Carbon Stars compared with an upgrade of the observed one
presented in \citealt{guanda} (dotted histogram; {\it courtesy of
R. Guandalini}).} \label{fig7}
\end{figure*}
In Fig. \ref{fig7} we report our theoretical LFCS (solid
histogram) and a revised version of the observed Galactic LFCS
presented by \citealt{guanda} (dotted histogram; {\it courtesy of
R. Guandalini}). The $y$-axis (fractional numbers) is defined as
the fraction of carbon stars in the bin of luminosity with respect
to the total number of carbon stars. The observed LFCS consists of
111 objects, whose estimate of absolute luminosity can be obtained
in two ways. Firstly, thanks to the use of Hipparcos parallaxes
recently revised \citep{hipparcos} and of IR data from the ISO and
MSX missions or, secondly, from the period-Luminosity relation for
Mira C-rich sources published in \citet{white} (see
\citealt{guanda} and \citealt{gubu} for the technical procedure
followed to construct the LFCS). The theoretical LFCS has been
obtained by assuming that the current AGB population of the
Galactic disk consists of stars born at different ages (and thus
with different masses) and with different metallicities. We
hypothesize that the disk formed 10 Gyr ago and, increasing the
age with time steps of 10$^5$ yr, we evaluate the number of AGB
stars currently laying on their TP-AGB during each time step. We
calculate the lower and upper mass on the AGB by interpolating the
physical inputs (such as the main sequence lifetime, the AGB
lifetime and the bolometric magnitudes along the AGB) on the grid
of computed models. The upper mass has been fixed to 3.0
M$_\odot$, while the minimum mass depends on the metallicity. In
order to cover the low mass tail of the LFCS, we calculate
additional models with a restricted nuclear network. At high
metallicity ($Z=2\times 10^{-2}$ and $Z=Z_\odot$) the minimum
initial mass for a star to become carbon rich is 1.5 $M_\odot$,
then decreases to 1.4 $M_\odot$ at $Z=10^{-2}$, to 1.3 $M_\odot$
at $Z=6\times 10^{-3}$, to 1.2 $M_\odot$ at $Z=3\times 10^{-3}$
and, finally, to 1.1 $M_\odot$ at $Z=10^{-3}$. We adopt a linear
decreasing Star Formation Rate (SFR)\footnote{We assume an initial
amount of gas converted to stars of 8 $M_\odot$ per yr and then we
apply a linear decrease down to a current value of 3 $M_\odot$ per
yr.} and the IMF proposed by \citet{salpe}. Moreover, we assume
a theoretical flat metallicity distribution (see below).\\
\begin{figure*}[tpb] \centering
\includegraphics[width=\textwidth]{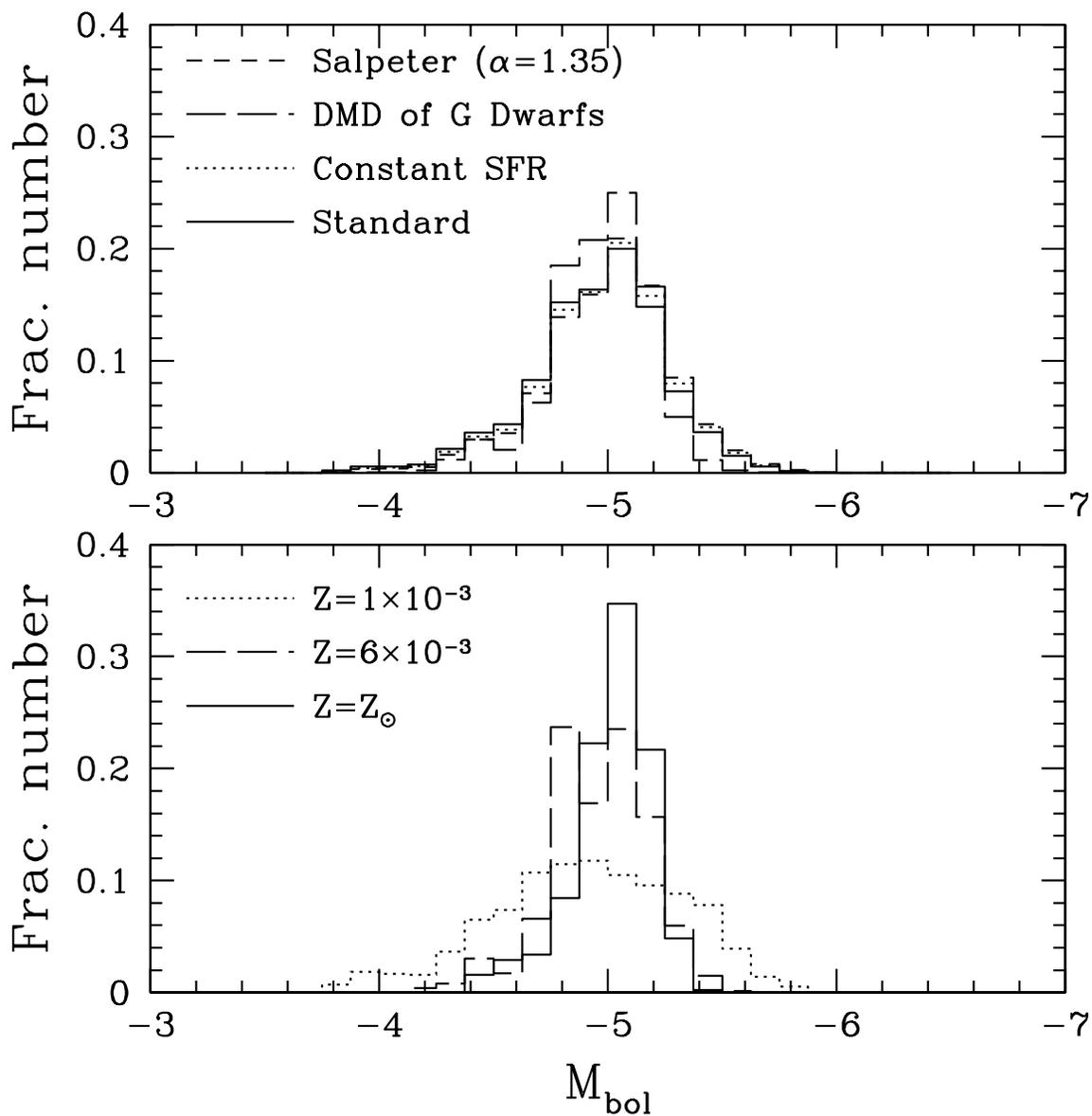}
\caption{Upper Panel: theoretical Galactic LFCS calculated with
different input parameters: Standard Case (solid curve), modified
Salpeter's IMF (short-dashed curve), constant SFR (dotted curve)
and metallicity distribution as observed in local G-Dwarfs
(long-dashed curve). Lower Panel: LFCS for selected key
metallicities: $Z=1\times 10^{-3}$ (dotted curve), $Z=6\times
10^{-3}$ (dashed curve) and $Z=Z_\odot$ (solid curve).}
\label{fig8}
\end{figure*}
Our curve matches the observed one, showing, within observational
errors, the same peak around $M_{\rm bol} \sim -5.0$. Note that,
unlike \citet{stan05}, we do not apply any normalization
coefficient to match the peak of the corresponding observations.
Moreover, we highlight that the choice of a different IMF, SFR or
metallicity distribution, does lead to very similar results. In
the upper panel of Fig.\ref{fig8} we report our standard
luminosity function (solid curve), a test case with a Salpeter's
IMF law with $\alpha$=1.35, a test case computed with a constant
SFR over the whole Galaxy evolution (dotted curve) and a test with
a metallicity distribution as derived from observations of local
G-dwarfs (Dual Metallicity Distribution, DMD; \citealt{rocha}). As
already stated, the differences between different cases are
negligible. Moreover, we evaluate the weight that each metallicity
has on the final luminosity function. In the lower panel of Fig.
\ref{fig8} we plot some LFCS corresponding to selected key
metallicities: $Z=1\times 10^{-3}$ (dotted curve), $Z=6\times
10^{-3}$ (dashed curve) and $Z=Z_\odot$ (solid curve). It emerges
that larger is the metallicity, thinner is the LFCS, being the
$M_{\rm bol}$ peak always around $-$5.0~. The flattest LFCS is
obtained at $Z=1\times 10^{-3}$, due to the decrease of the
stellar mass for the occurrence of TDU (which accounts for the
faintest stars) and to the large increase of the final core masses
for stars with initial masses $M\ge 2.5 M_\odot$ (which accounts
for the brightest stars).
\begin{figure*}[tpb] \centering
\includegraphics[width=\textwidth]{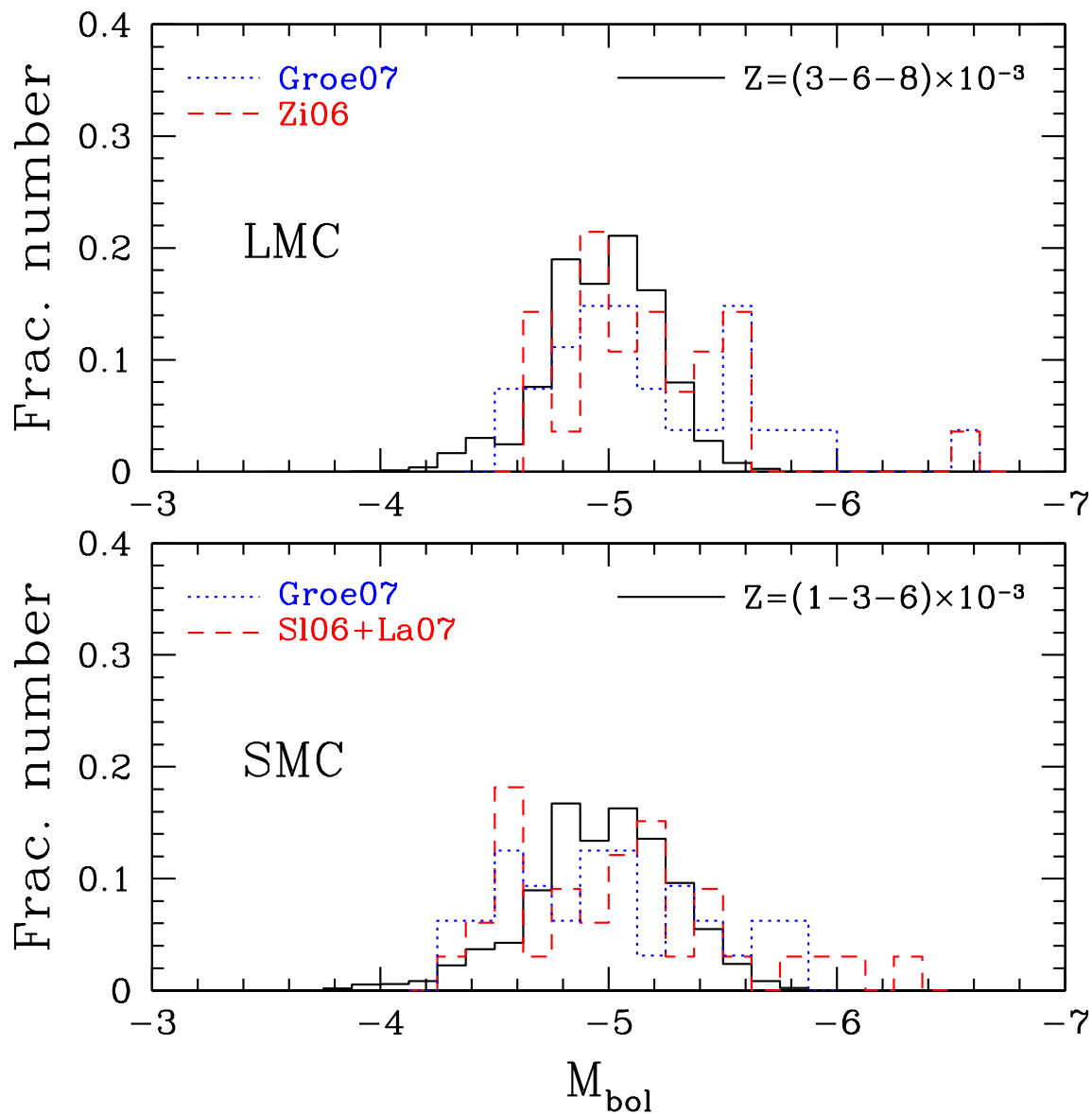}
\caption{Upper Panel: theoretical and observational LFCS for the
LMC system (Groe07: \citealt{groe07}; Zi06: \citealt{zil}); Lower
Panel: theoretical and observational LFCS for the SMC system
(Sl06: \citealt{sloan}; La07: \citealt{lagadec}). See text for
details.} \label{fig8_bis}
\end{figure*}

More stringent constraints would come from the comparison of
theoretical LFCS with the ones derived in the Magellanic Clouds
(MCs). This basically derives from a more precise determination of
MCs distances with respect to Galactic carbon stars.
Notwithstanding, we must remind that the photometric measurements
of C-rich stars belonging to MCs are often incomplete, since they
lack of the contribution from the thermal IR band ($\lambda > 2
\mu m$). As a matter of fact, these objects emit the majority of
their flux ($\sim$90\%) at these wavelengths, while the most
widely adopted data used to built up the LFCS were based on
optical (visible or near-IR data) by \citet{west} and \citet{rebe}
for SMC, or \citet{froca} for LMC. Recently, \citet{zil},
\citet{sloan}, \citet{groe07} and \citet{lagadec} report the
luminosities of a sample of about 60 C-stars belonging to MCs, as
derived by combining ground-based observations in the near-IR
(J,H,K and L) with {\it Spitzer Space Telescope} spectroscopic
observations. In the upper panel of Fig. \ref{fig8_bis} we compare
the LMC LFCS as reported by \citet{groe07} (dotted curve) and
\citet{zil} (dashed curve). Although the two observational works
share the same data set, the resulting Luminosity Functions appear
rather different. Nevertheless, they both peak at M$_{\rm
bol}=-5$, thus in good agreement with our theoretical LFCS. This
latter curve have been constructed by assuming a constant SFR and
3 equally weighted metallicities ($Z=3\times 10^{-3}$, $Z=6\times
10^{-3}$ and $Z=8\times 10^{-3}$). Similar conclusions can be
drawn for SMC (see the lower panel of Fig. \ref{fig8_bis}). In
that case the observed distributions are from \citet{groe07}
(dotted curve) and a combined sample from \citet{sloan} and
\citet{lagadec} (dashed curve). For SMC, the observational LFCS
shows a larger spread, but the average M$_{\rm bol}$ is similar to
the LMC one. The lower SMC mean metallicity forces us to use
models with different $Z$ to construct our theoretical LFCS
($Z=1\times 10^{-3}$, $Z=3\times 10^{-3}$ and $Z=6\times
10^{-3}$). As already explained, this makes our distribution
flatter and, thus, in quite
good agreement with observations.\\
The statistical samples used for the observed CSLFs reported in
Fig. \ref{fig8_bis} are quite poor, but we are confident that
future developments of ground and space-based IR-facilities would
greatly improve the observational constraints to AGB theoretical
models.

\subsection{{\it s}-process indexes}\label{hsls}

In this section we compare the predicted spectroscopic indexes
[hs/ls] and [Pb/hs] to the observed composition of various {\it
s}-enhanced stars. Among these stars, we should distinguish
between AGB stars ({\it i.e.} those presently undergoing thermal
pulses and TDU episodes), post-AGB stars and less-evolved stars
(dwarfs or giants) belonging to binary systems. In the latter
case, the {\it s}-enhancement is likely due to the pollution
caused by the intense wind of an AGB companion (hereinafter
extrinsic {\it s}-enhanced
stars).\\
In principle, such a comparison requires the knowledge of the
evolutionary status of the star, the initial mass and the
metallicity. In the case of extrinsic {\it s}-enhanced stars, in
addition to the initial masses and the metallicity of both the
binary components, we should know the evolutionary status of the
presently observed secondary star, its age at the epoch of the
pollution and the total mass of chemically enriched material
deposited on its surface (\emph{Bisterzo et al., submitted}).
Nevertheless, according to our low-mass-AGB models,  the two
spectroscopic indexes [hs/ls] and [Pb/hs] should roughly attain
the final value after a few third dredge-up episodes. In other
words, if the {\it s}-process enhancement is sufficiently large,
the knowledge of the precise evolutionary status is no longer a
necessary condition. In practice, if the {\it s}-enhancement has
been caused by the radiative $^{13}$C burning, as we have shown to
occur in a low mass AGB stars, the [hs/ls] and the [Pb/hs] should
only depend on the stellar metallicity and, eventually, on its
mass.

In the following analysis we consider {\it s}-enhanced stars
spectroscopically classified as: S stars (O-rich AGB stars), N
type C stars (C-rich AGB stars) post-AGB (either O- and C-rich),
Ba and/or CH stars (C-rich extrinsic {\it s}-enhanced stars). We
also consider a few SC stars, even though their evolutionary
status and origin are unknown.
\begin{figure*}[tpb]
\centering
\includegraphics[width=\textwidth]{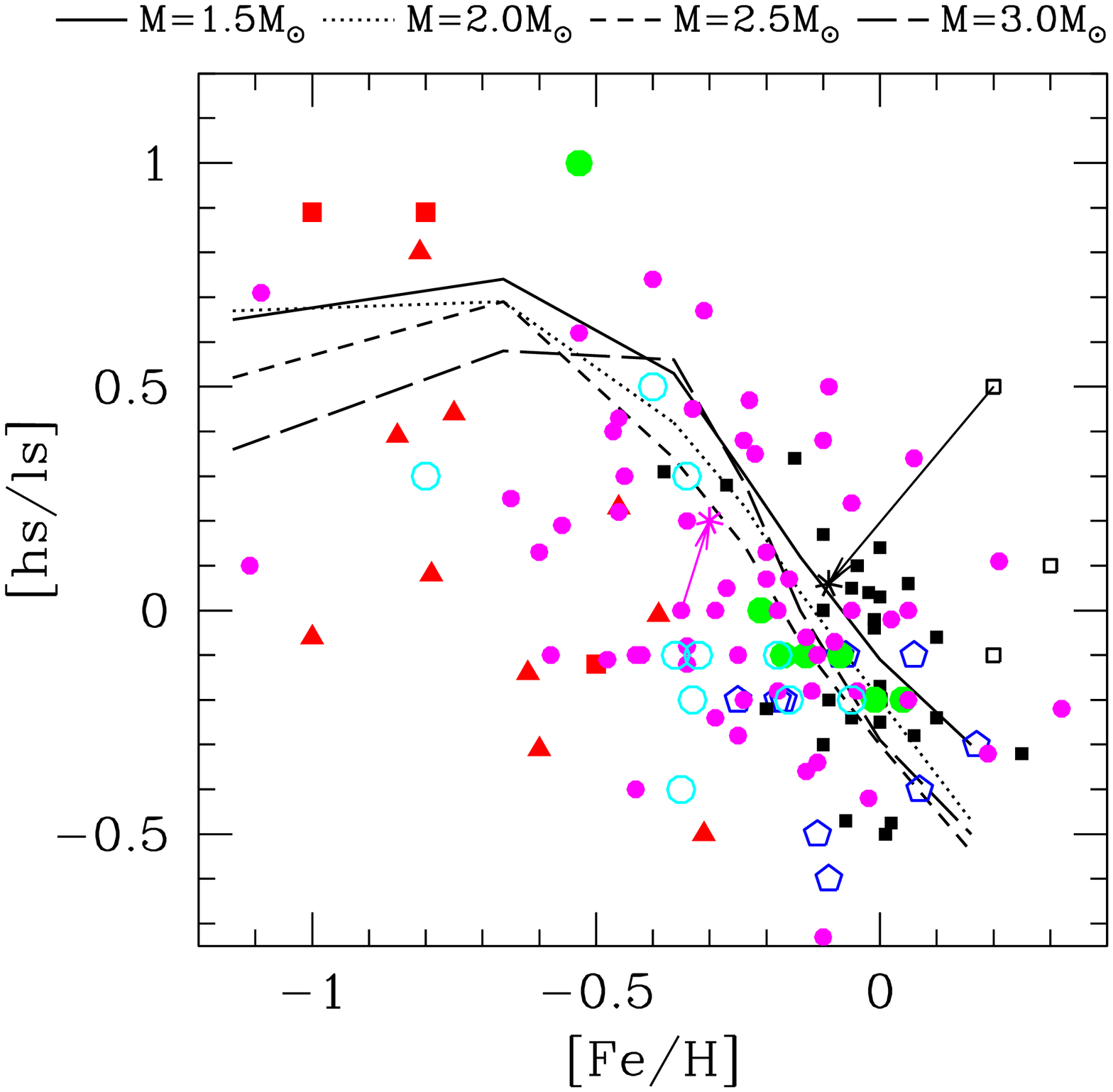}
\caption{Our theoretical [hs/ls] curves compared with
observational data. We report: intrinsic O-stars (open blue
pentagon: \citealt{smith85,smith86,smith90}); intrinsic SC-stars
(little open dark squares: \citealt{ab98}); intrinsic C-stars
(little filled dark squares: \citealt{ab02,zamora09}); post-AGB
stars (filled red triangles: \citealt{vanwi,reddy,rey04});
extrinsic O-rich stars (large filled green circles:
\citealt{smith85,smith86,smith90}), Ba-stars (little filled
magenta circles:
\citealt{smith84a,smith84b,kova85,tom86,smith87,smith90,pereira98,ab06,smilia,liu});
CH-stars (large open cyan circles:
\citealt{vanture,smith93,zacs,pereira2011}). Correction for the SC
star RR Her \citep{zamo2} and for the Ba-star HR107 (this work)
are also reported.} \label{fig9}
\end{figure*}
Then, in Fig. \ref{fig9} we compare, as a function of the
metallicity,  our final surface [hs/ls] with the corresponding
measured values, for a rather large sample of stars. In the
caption we report the references corresponding to observational
data. Although a similar trend with the metallicity may be
recognized, the rather large spread of the observed [hs/ls] at any
metallicity cannot be explained through a change of the stellar
mass.

The comprehension of the origin of this spread is a hard task.
Note that the typical error associated to the derivation of the
heavy elements abundances for these {\it s}-enhanced stars is of
the order of $\pm 0.1$ dex, even if a more conservative evaluation
might suggest up to $\pm 0.3$ dex. Such an error, however, only
takes into account the uncertainty introduced in the abundance
analysis by the unknown stellar parameters (effective temperature,
gravity, metallicity). The systematic uncertainty associated to
the particular physical prescriptions used to model the stellar
atmosphere, such as the deviations from the local thermodynamics
equilibrium (LTE) or from the spherical symmetry (1D versus 3D
models), is often neglected. As a matter of fact, the use of
different procedures for the abundance analysis and/or model
atmospheres might lead to different stellar parameters and, hence,
to different results. A couple of examples may illustrate this
problem. We found at least two stars in our sample for which a
revision of the abundance analysis is possible. The SC-type star
RR Her was analyzed by \citet{ab98} by using atmosphere models
from Alexander et al. (unpublished). For this star, they obtained
[Fe/H]=0.2 and an [hs/ls]=0.5. Recently, \citet{zamo2} reanalyzed
this star by using the state-of-the-art of atmosphere models for
C-stars \citep{gusta08}, together with updated molecular and
atomic line lists. The new derived [Fe/H] is -0.09 and the [hs/ls]
ratio is 0.06. As a result, the revised [hs/ls] nicely fits the
theoretical prediction (see Fig. \ref{fig9}). Moreover, we were
also able to re-analyze the Ba-star HR 107, previously studied by
\citet{ab06}. We take advantage of a very high-resolution (R$\sim
200000$) HARPS visual spectrum available in the ESO archive. Then,
by adopting the same atmosphere parameters as in \citet{ab06}, we
derive a [hs/ls]=0.2 to be compared with the [hs/ls]=0 obtained by
the authors. A more detailed description of the tools used for the
revised analysis may be found in \citet{ab02}, \citet{dela06} and
\citet{zamora09}.

Lead measurements are also available for a sub-sample of the
observed stars \citep{ab06}. The comparison with the theoretical
[Pb/hs] index is shown in Fig. \ref{fig10}. As for the [hs/ls]
ratio, the general trend of the observed data appears in
reasonable agreement with the theoretical expectations. Also in
this case, however, the observed spread cannot be ascribed to a
variation of the stellar mass within the range $1.5-3.0 M_\odot$.
We have also revised the lead abundance of the Ba stars HR 107.
From the HARPS high-resolution spectrum we have obtained an upper
limit [Pb/hs]=0.05, which is definitely smaller than the 0.34
quoted by \citet{ab06}. This confirms the need of more accurate
evaluation of the systematic uncertainties affecting the abundance
analysis.

Although it is possible that the observed spread may be (totally
or in part) ascribed to a large observational uncertainty, we
mention possible alternative theoretical scenarios, which may
result in a certain spread of the spectroscopic indexes [hs/ls]
and [Pb/hs]. As discussed in Section \ref{nucleo}, these indexes
basically depend on 3 quantities, namely: neutrons, poisons and
seeds. A change of the amount of neutrons would imply a change of
the ({\it effective}) $^{13}$C in the pocket. In the post-process
calculations performed by \citealt{ga98} (see also
\citealt{bu01,pigna,husti,bi11}), the amount of $^{13}$C within
the pocket (and accordingly the amount of $^{14}$N)  has been
varied parametrically in order to match the observed chemical
pattern. In this way, since the extension of the pocket and, in
turn, the amount of Fe seeds are unchanged, the neutrons-to-seed
ratio can be freely varied, in order to reproduce the observed
spread of the spectroscopic indexes. Such a procedure is justified
by the uncertain theoretical description of the hydrodynamical
process which determines the proton profile in the transition zone
between the fully-convective envelope and the radiative
H-exhausted core during a TDU episode. As a matter of fact, in our
model such a profile is the result of the assumed exponential
decay of the average mixing velocity at the inner border of the
convective envelope. A spread of such an average description may
eventually imply a spread in the resulting profile. Note, however,
that even a small variation (positive or negative) of the $\beta$
parameter determining the shape of the convective velocities
within the convective-radiative transition region strongly
depletes the resulting {\it s}-process nucleosynthesis, as
demonstrated in Section \ref{uncer} (see also Paper I). This would
lead to definitely lower final surface heavy elements abundances,
thus in contrast with observations. Moreover, the C-star
luminosity function would be altered (see Section \ref{cstarlum}).
Such an occurrence provides an independent check of the adopted
parametrization. As we have already shown, the predicted and the
observed LF of the Galactic C-stars are in good agreement.

\begin{figure*}[tpb] \centering
\includegraphics[width=\textwidth]{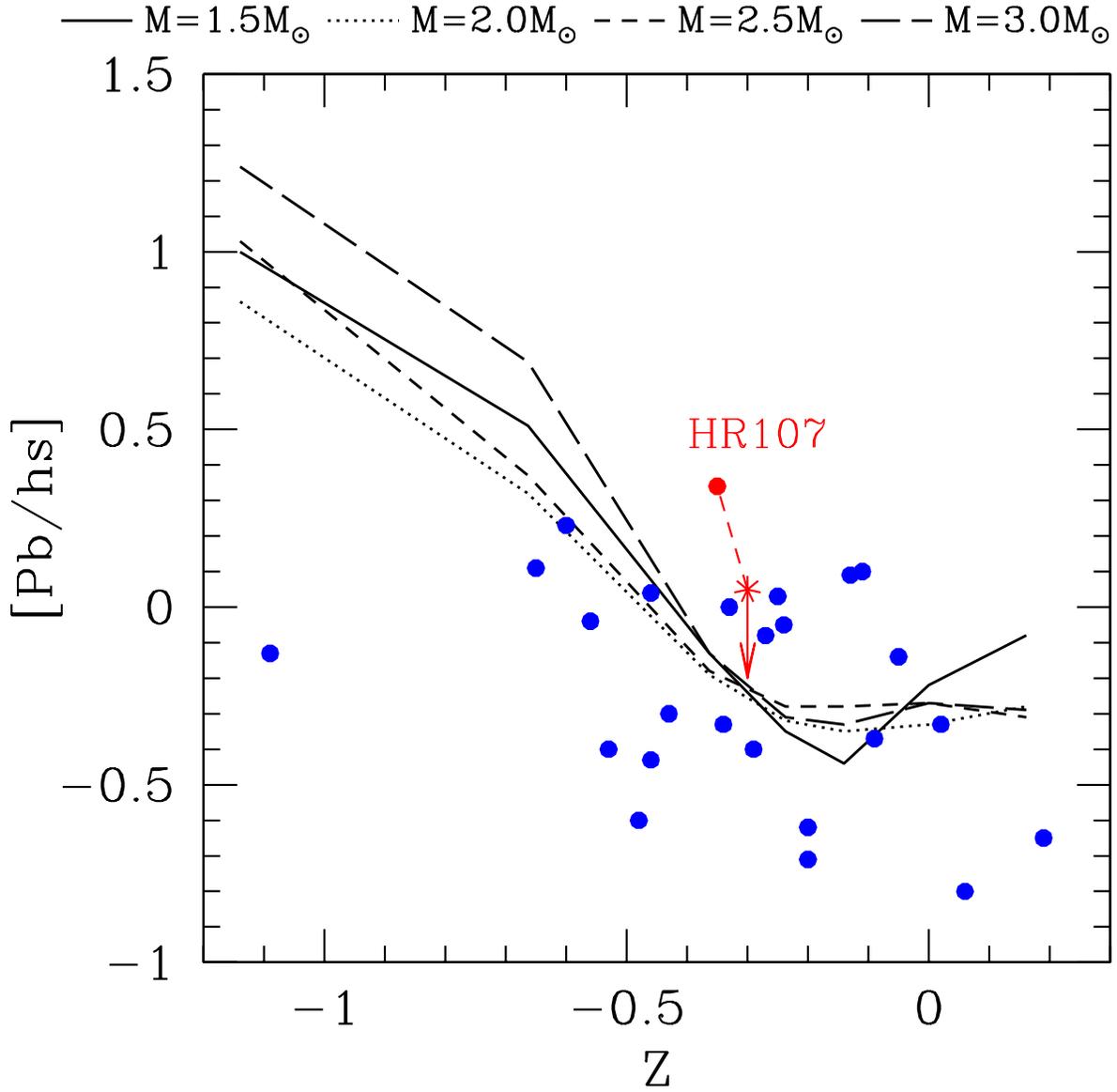}
\caption{Theoretical [Pb/hs] curves compared with the same ratios
as observed in a sample of Galactic barium stars \citep{ab06}. The
new upper limit derived for HR107 (this work) is also reported.}
\label{fig10}
\end{figure*}

Alternative scenarios may eventually account for the observed
spread of the spectroscopic indexes, leaving the LFCS unchanged.
Mixing induced by rotation has been invoked to explain the
formation of the $^{13}$C pocket \citep{la99}. The common
conclusion of all these attempts is that the resulting $^{13}$C
pocket is too small for a sizeable {\it s}-process
nucleosynthesis. This likely occurs because, during TDU episodes,
the timescale for the formation of a proton profile (and the
subsequent $^{13}$C pocket) in the upper layers of the
He-intershell is too short compared to the timescale
characterizing rotational-induced instabilities. Nevertheless,
these instabilities might act during the longer time elapsed
between the formation of the proton profile (maximum depth of the
TDU) and the beginning of the {\it s}-process nucleosynthesis
\citep{he03,si04}. In low-mass AGB stars of nearly solar
composition, such a period may be as long as a few $10^4$ yr. If
enough total angular momentum is conserved at the beginning of the
TP-AGB phase, the Eddington-Sweet instability, in particular, may
have enough time to smear the $^{13}$C (and the $^{14}$N) pocket,
thus reducing the local neutrons-to-seeds ratio. A quantitative
analysis of this hypothesis is beyond the purpose of the present
work and will be addressed in a forthcoming paper.

Finally, let us note that when the {\it s}-process nucleosynthesis
takes place at the base of a quite extended convective zone, the
seeds consumption is counterbalanced by the rapid mixing of fresh
material from outside. On the contrary, due to the rapid
destruction compared to the production, the amount of neutrons is
maintained at a local equilibrium value, which depends on the
temperature. As a result, a relatively low neutrons-to-seed ratio
is maintained compared to the case of radiative burning, thus
favoring the accumulation of light-{\it s} elements. In massive
AGB stars, where the main neutron source is the
$^{22}$Ne$(\alpha,n)^{25}$Mg reaction, which is active at the base
of the convective shell, we expect an overproduction of ls. The
case of the 3 $M_\odot$ model with $Z=0.001$, where a substantial
activation of the $^{22}$Ne convective burning is found, shows in
fact a particularly low [hs/ls] compared to lower mass models with
the same metallicity.

\section{Conclusions}\label{concl}

We presented FRUITY, an on-line database entirely dedicated to the
nucleosynthesis occurring in AGB stars. Its interface is hosted on
the web-pages of the Teramo Observatory (INAF-OACTe) and is
accessible at the internet address
http://www.oa-teramo.inaf.it/fruity. In this paper we discussed
and analyzed the physical and chemical evolution of the first set
of models included in the database, currently focused on low mass
AGB stars (1.5 $\le M/M_{\odot} \le$ 3.0) with metallicities in
the range $1\times 10^{-3} \le Z \le 2\times 10^{-2}$. In the
future, we intend to extend the grid of models by increasing the
mass (1.0 $< M/M_\odot < $ 7.0) and metallicity ($-3.2 <$ [Fe/H]$
<-1.2$) ranges. Our calculations include a full nuclear network,
from H to Bi, with 700 isotopes linked by more than 1000 nuclear
processes. Thus, we followed in detail the nucleosynthesis of both
light and heavy ({\it s}-process) elements. On the web interface,
models selection can be operated choosing the desired mass ($M$)
and metallicity ($Z$) values, while properties of nuclides
(elements, isotopes, {\it s}-process spectroscopic indexes and
yields) can be retrieved for all masses and atomic numbers or only
for a sub-sample (selecting an appropriate pair of A/Z values).
Moreover, the surface distributions after each TDU or just the
final one can be selected for each model. Elemental abundances are
given in the usual spectroscopic notation, while mass fractions
are returned for the isotopes. Finally, net yields are given in
solar mass units.

We analyzed the effects deriving from the activation of both the
neutron sources in our low mass AGB models, the
$^{13}$C($\alpha$,n)$^{16}$O and the
$^{22}$Ne($\alpha$,n)$^{25}$Mg reactions, highlighting their
contribution at different metallicities and for different initial
masses. We evaluated the sensitivity of our model predictions to
the adopted input parameters, with particular emphasis on the
treatment of convection (the $\alpha$ parameter of the mixing
length and the $\beta$ parameter determining the shape of the
exponentially decreasing velocity profile at the inner border of
the convective envelope) and the mass loss law. We compared our
model with the observational counterparts, from both a physical
(luminosity function of carbon stars) and chemical ({\it
s}-process spectroscopic indexes at different metallicities) point
of view. Our theoretical luminosity function of carbon stars well
fits the observational one, recently derived with the
state-of-the-art set of distances (HIPPARCOS) and spectra (ISO).
If recent SPITZER data are taken into account, a good agreement
was also found for the Magellanic Clouds. In the future this
comparison could be improved for the C-star population of nearby
galaxies, as soon as a larger number of IR data is available. A
comparison between our theoretical curves and spectroscopic data
showed that, at fixed metallicity, our models cannot account for
the {\it s}-process indexes spread identified by observations (in
particular the [hs/ls] ratio). In the framework of our current
modelling, we excluded that such a spread could be ascribed to the
different initial masses of AGB models. We speculated on possible
solutions from both an observational and theoretical point of
view. In this latter case, we suggested that rotation may (at
least partially) solve the current disagreement.

\acknowledgements {}

Part of this work was supported by the Spanish grants
AYA2008-04211-C02-02 and FPA2008-03908 from the MEC and the
Italian Grant FIRB 2008 "FUTURO IN RICERCA" C81J10000020001. We
deeply thank R. Guandalini for providing unpublished data. We
thank the anonymous referee for a detailed reading of this paper and for
helpful comments and suggestions.

\bibliographystyle{99}

\newpage

\begin{deluxetable}{ccccccc}
\tablecolumns{7} \tablewidth{0pc} \tablecaption{The C-enhancement
factors for the H-rich opacity tables at various metallicities,
defined as the ratio between the $^{12}$C abundance used to
calculate the opacity table and the initial $^{12}$C.}
\tablehead{\colhead{$Z$} & \colhead{[Fe/H]} &
\multicolumn{5}{c}{enhancement factor ($f_i$)}}
\startdata & & $f_1$ & $f_2$ & $f_3$ & $f_4$ & $f_5$ \\
\cline{1-7}
0.020 & 0.16  & 1 & 1.5 & 1.8  & 2.2 & 5\\
0.014 & 0     & 1 & 1.5 & 1.8  & 2.2 & 4\\
0.010 & -0.14 & 1 & 1.8 & 2.2  & 5 & 10\\
0.008 & -0.24 & 1 & 1.8 & 2.2  & 5 & 10\\
0.006 & -0.36 & 1 & 2   & 5    & 10  & 50\\
0.003 & -0.66 & 1 & 2   & 5    & 10  & 50\\
0.001 & -1.14 & 1 & 5 & 10 & 50  & 200\\
\enddata
 \label{tab1}
\end{deluxetable}

\begin{deluxetable}{ccccccccccccc}
\rotate
 \tablecolumns{9} \tablewidth{0pc}
\tablecaption{Fundamental properties of the pre-AGB phase.}
\tablehead{ \colhead{$M$ ($M_\odot$)} & \colhead{$\Delta t_{\rm
H}$ (Myr)} & \colhead{$M_{\rm H}^{cc}$ ($M_\odot$)} &
\colhead{He$_{surf}^{FDU}$} & \colhead{log $L_{tip}^{RGB}$} &
\colhead{$M_{\rm He}^{1}$ ($M_\odot$)} & \colhead{$\Delta t_{\rm
He}$ (Myr)} & \colhead{$M_{\rm He}^{2}$ ($M_\odot$)} &
\colhead{$M_{\rm He}^{3}$ ($M_\odot$)}}\startdata
\cutinhead{\textbf{$Z$=0.02 \ $Y$=0.269}}
1.5  & 2839 & 0.122 & 0.289 & 3.428 & 0.480 & 116 & 0.525 & 0.547 \\
2.0  & 1159 & 0.270 & 0.280 & 3.376 & 0.471 & 122 & 0.527 & 0.549 \\
2.5  & 592  & 0.444 & 0.282 & 2.421 & 0.314 & 302 & 0.489 & 0.523 \\
3.0  & 346  & 0.594 & 0.285 & 2.497 & 0.369 & 168 & 0.536 & 0.555 \\
\cutinhead{\textbf{$Z$=0.014 \ $Y$=0.269}}
1.5  & 2310 & 0.131 & 0.286 & 3.421 & 0.481 & 109 & 0.531 & 0.550 \\
2.0  & 970  & 0.293 & 0.280 & 3.333 & 0.465 & 119 & 0.531 & 0.548 \\
2.5  & 503  & 0.469 & 0.282 & 2.448 & 0.315 & 259 & 0.506 & 0.530 \\
3.0  & 299  & 0.628 & 0.285 & 2.523 & 0.375 & 138 & 0.562 & 0.581 \\
\cutinhead{\textbf{$Z$=0.01 \ $Y$=0.265}}
1.5  & 2162 & 0.136 & 0.284 & 3.412 & 0.483 & 105 & 0.532 & 0.549 \\
2.0  & 913  & 0.301 & 0.277 & 3.322 & 0.466 & 117 & 0.531 & 0.551 \\
2.5  & 479  & 0.496 & 0.279 & 2.443 & 0.317 & 242 & 0.506 & 0.530 \\
3.0  & 288  & 0.646 & 0.281 & 2.523 & 0.378 & 125 & 0.577 & 0.597 \\
\cutinhead{\textbf{$Z$=0.008 \ $Y$=0.265}}
1.5  & 1983 & 0.141 & 0.284 & 3.403 & 0.483 & 103 & 0.535 & 0.552 \\
2.0  & 845  & 0.310 & 0.277 & 3.291 & 0.463 & 116 & 0.533 & 0.551 \\
2.5  & 447  & 0.483 & 0.279 & 2.431 & 0.319 & 218 & 0.517 & 0.538 \\
3.0  & 271  & 0.663 & 0.280 & 2.527 & 0.382 & 110 & 0.607 & 0.622 \\
\cutinhead{\textbf{$Z$=0.006 \ $Y$=0.260}}
1.5  & 1851 & 0.147 & 0.280 & 3.392 & 0.484 & 100 & 0.537 & 0.556 \\
2.0  & 810  & 0.327 & 0.273 & 3.268 & 0.462 & 114 & 0.539 & 0.555 \\
2.5  & 430  & 0.496 & 0.274 & 2.419 & 0.317 & 199 & 0.531 & 0.549 \\
3.0  & 264  & 0.664 & 0.273 & 2.524 & 0.383 & 97.5& 0.642 & 0.656 \\
\cutinhead{\textbf{$Z$=0.003 \ $Y$=0.260}}
1.5  & 1607 & 0.169 & 0.280 & 3.358 & 0.484 & 95.7& 0.545 & 0.562 \\
2.0  & 712  & 0.336 & 0.274 & 3.176 & 0.451 & 114 & 0.550 & 0.569 \\
2.5  & 382  & 0.509 & 0.273 & 2.392 & 0.324 & 151 & 0.595 & 0.613 \\
3.0  & 238  & 0.702 & 0.266 & 2.515 & 0.383 & 79.6& 0.717 & 0.724 \\
\cutinhead{\textbf{$Z$=0.001 \ $Y$=0.245}}
1.5  & 1551 & 0.189 & 0.268 & 3.292 & 0.484 & 91.0& 0.553 & 0.574 \\
2.0  & 679  & 0.365 & 0.263 & 3.108 & 0.452 & 102 & 0.588 & 0.598 \\
2.5  & 373  & 0.538 & 0.255 & 2.315 & 0.329 & 124 & 0.660 & 0.669 \\
3.0  & 236  & 0.741 & 0.247 & 2.464 & 0.397 & 70.1& 0.775 & 0.779 \\
\enddata \label{tab2}
\end{deluxetable}

\begin{deluxetable}{lcccccccccccc}
\rotate \tablecolumns{13} \tablewidth{0pc} \tablecaption{Evolution
of selected physical parameters of the models at $Z=2\times
10^{-2}$.} \tablehead{ \colhead{$n_{\rm TP}^a$} & \colhead{$M_{\rm
Tot}^{b}$} & \colhead{$M_{\rm H}^{b}$} & \colhead{$\Delta M_{\rm
TDU}^{b}$} & \colhead{$\Delta M_{\rm CZ}^{b}$} & \colhead{$\Delta
M_{\rm H}^{b}$} & \colhead{$r$} & \colhead{$\lambda$}
&\colhead{$\Delta t_{ip}^{c}$} & \colhead{$T_{MAX}^{d}$} &
\colhead{Z$_{surf}$} & \colhead{C/O$_{\rm surf}$} &
\colhead{$n_c$}}\startdata \cutinhead{\textbf{$M$=1.5 $M_\odot$ }}
-4 & 1.32 & 0.542 & 0.00E+00 & 3.14E-02 & 2.20E-03 & 9.92E-01 & 0.00E+00 &  11.3 & 2.06 & 2.02E-02 & 0.34         &  \nodata \\
-3$^{(*)}$ & 1.31 & 0.547 & 0.00E+00 & 3.22E-02 & 4.40E-03 & 9.04E-01 & 0.00E+00 &  15.5 & 2.24 & 2.02E-02 & 0.34 &  \nodata \\
-2 & 1.30 & 0.552 & 0.00E+00 & 3.11E-02 & 5.60E-03 & 8.54E-01 & 0.00E+00 &  17.1 & 2.34 & 2.02E-02 & 0.34         &  \nodata \\
-1 & 1.28 & 0.559 & 0.00E+00 & 3.04E-02 & 6.80E-03 & 8.02E-01 & 0.00E+00 &  17.6 & 2.45 & 2.02E-02 & 0.34         &  \nodata \\
1  & 1.26 & 0.566 & 6.00E-04 & 2.89E-02 & 7.10E-03 & 7.65E-01 & 8.45E-02 &  17.1 & 2.52 & 2.03E-02 & 0.36         & 5 \\
2  & 1.23 & 0.574 & 1.30E-03 & 2.75E-02 & 8.20E-03 & 7.11E-01 & 1.59E-01 &  16.2 & 2.58 & 2.07E-02 & 0.42         & 27 \\
3  & 1.20 & 0.581 & 2.00E-03 & 2.62E-02 & 8.80E-03 & 6.75E-01 & 2.27E-01 &  15.2 & 2.64 & 2.14E-02 & 0.53         & 34 \\
4  & 1.16 & 0.588 & 2.30E-03 & 2.49E-02 & 9.20E-03 & 6.42E-01 & 2.50E-01 &  14.1 & 2.68 & 2.20E-02 & 0.67         & 36 \\
5  & 1.10 & 0.595 & 7.00E-04 & 2.36E-02 & 9.20E-03 & 6.20E-01 & 7.61E-02 &  13.0 & 2.72 & 2.28E-02 & 0.71         & 28\\
\cline{1-13} & \multicolumn{3}{r}{$M_{\rm TDU}^{\rm Tot}$ =
\textbf{6.90E-03 $M_\odot$}} & & & & \multicolumn{3}{r}{$\tau_C$ =
\textbf{0 yr}}& & &
\\
\\
\cutinhead{\textbf{$M$=2.0 $M_\odot$ }}
-3 & 1.89 & 0.543 & 0.00E+00 & 3.18E-02 & 3.10E-03 & 9.31E-01 & 0.00E+00 &  12.5 & 2.12 & 2.02E-02 & 0.31 & \nodata        \\
-2$^{(*)}$ & 1.89 & 0.549 & 0.00E+00 & 3.18E-02 & 5.40E-03 & 8.66E-01 & 0.00E+00 &  16.1 & 2.26 & 2.02E-02& 0.31 &           \nodata \\
-1 & 1.88 & 0.555 & 0.00E+00 & 3.12E-02 & 6.20E-03 & 8.30E-01 & 0.00E+00 &  18.0 & 2.39 & 2.02E-02 & 0.31 & \nodata        \\
1  & 1.86 & 0.562 & 8.00E-04 & 3.00E-02 & 6.80E-03 & 7.85E-01 & 1.18E-01 &  18.1 & 2.48 & 2.03E-02 & 0.32 & 17 \\
2  & 1.85 & 0.569 & 1.70E-03 & 2.87E-02 & 8.30E-03 & 7.21E-01 & 2.05E-01 &  17.5 & 2.56 & 2.06E-02 & 0.37 & 33 \\
3  & 1.83 & 0.577 & 3.10E-03 & 2.75E-02 & 9.10E-03 & 6.82E-01 & 3.41E-01 &  16.8 & 2.63 & 2.12E-02 & 0.45 & 36 \\
4  & 1.80 & 0.583 & 4.00E-03 & 2.65E-02 & 1.00E-02 & 6.32E-01 & 4.00E-01 &  16.1 & 2.70 & 2.20E-02 & 0.57 & 31 \\
5  & 1.76 & 0.590 & 4.60E-03 & 2.55E-02 & 1.05E-02 & 5.94E-01 & 4.38E-01 &  15.4 & 2.75 & 2.30E-02 & 0.70 & 29 \\
6  & 1.70 & 0.596 & 4.90E-03 & 2.44E-02 & 1.08E-02 & 5.69E-01 & 4.54E-01 &  14.5 & 2.79 & 2.41E-02 & 0.84 & 29 \\
7  & 1.61 & 0.602 & 4.90E-03 & 2.33E-02 & 1.08E-02 & 5.44E-01 & 4.54E-01 &  13.5 & 2.83 & 2.51E-02 & 0.99 & 28 \\
8  & 1.49 & 0.608 & 4.70E-03 & 2.22E-02 & 1.07E-02 & 5.25E-01 & 4.39E-01 &  12.4 & 2.85 & 2.63E-02 & 1.15 & 26 \\
9  & 1.34 & 0.614 & 4.50E-03 & 2.11E-02 & 1.04E-02 & 5.17E-01 & 4.33E-01 &  11.4 & 2.87 & 2.76E-02 & 1.33 & 20 \\
10 & 1.17 & 0.619 & 1.15E-03 & 2.00E-02 & 9.60E-03 & 5.14E-01 & 1.20E-01 &  10.2 & 2.88 & 2.76E-02 & 1.33 &   19 \\
\cline{1-13} & \multicolumn{3}{r}{$M_{\rm TDU}^{\rm Tot}$ =
\textbf{3.44E-02 $M_\odot$}} & & & & \multicolumn{3}{r}{$\tau_C$ =
\textbf{3.39E+05 yr}}& & &
\\
\\
\cutinhead{\textbf{$M$=2.5 $M_\odot$ }}
-1$^{(*)}$ & 2.47 & 0.523 & 0.00E+00 & 3.76E-02 & 6.70E-03 & 8.52E-01 & 0.00E+00 &  23.0 & 2.25 & 2.02E-02 & 0.31 &  \nodata \\
1  & 2.46 & 0.531 & 8.00E-04 & 3.70E-02 & 7.30E-03 & 7.50E-01 & 8.16E-02 &  36.1 & 2.42 & 2.04E-02 & 0.34         &  5 \\
2  & 2.45 & 0.540 & 1.50E-03 & 3.39E-02 & 9.00E-03 & 7.48E-01 & 2.05E-01 &  25.8 & 2.44 & 2.05E-02 & 0.36         &  16 \\
3  & 2.44 & 0.548 & 2.80E-03 & 3.27E-02 & 9.30E-03 & 7.29E-01 & 3.01E-01 &  23.4 & 2.52 & 2.08E-02 & 0.40         &  30 \\
4  & 2.43 & 0.555 & 4.00E-03 & 3.15E-02 & 1.01E-02 & 6.91E-01 & 3.96E-01 &  21.9 & 2.59 & 2.13E-02 & 0.48         &  35 \\
5  & 2.41 & 0.562 & 5.20E-03 & 3.04E-02 & 1.09E-02 & 6.53E-01 & 4.77E-01 &  20.7 & 2.65 & 2.20E-02 & 0.57         &  30 \\
6  & 2.39 & 0.568 & 6.00E-03 & 2.94E-02 & 1.16E-02 & 6.15E-01 & 5.17E-01 &  19.7 & 2.70 & 2.28E-02 & 0.68         &  27 \\
7  & 2.37 & 0.574 & 6.60E-03 & 2.84E-02 & 1.21E-02 & 5.86E-01 & 5.45E-01 &  18.6 & 2.75 & 2.37E-02 & 0.80         &  26 \\
8  & 2.34 & 0.580 & 6.70E-03 & 2.73E-02 & 1.23E-02 & 5.61E-01 & 5.45E-01 &  17.5 & 2.78 & 2.45E-02 & 0.91         &  24 \\
9  & 2.30 & 0.585 & 6.50E-03 & 2.61E-02 & 1.22E-02 & 5.43E-01 & 5.33E-01 &  16.3 & 2.81 & 2.53E-02 & 1.02         &  23 \\
10 & 2.24 & 0.591 & 6.30E-03 & 2.50E-02 & 1.20E-02 & 5.30E-01 & 5.25E-01 &  15.1 & 2.84 & 2.61E-02 & 1.13         &  22 \\
11 & 2.16 & 0.596 & 6.10E-03 & 2.38E-02 & 1.18E-02 & 5.16E-01 & 5.17E-01 &  14.0 & 2.85 & 2.69E-02 & 1.24         &  22 \\
12 & 2.06 & 0.602 & 5.90E-03 & 2.28E-02 & 1.15E-02 & 5.07E-01 & 5.13E-01 &  13.0 & 2.87 & 2.77E-02 & 1.36         &  22 \\
13 & 1.92 & 0.607 & 5.70E-03 & 2.19E-02 & 1.12E-02 & 4.99E-01 & 4.09E-01 &  12.0 & 2.89 & 2.85E-02 & 1.48         &  21 \\
14 & 1.76 & 0.612 & 5.30E-03 & 2.09E-02 & 1.09E-02 & 4.90E-01 & 4.86E-01 &  11.2 & 2.91 & 2.95E-02 & 1.60         &  21 \\
15 & 1.57 & 0.617 & 4.90E-03 & 2.00E-02 & 1.04E-02 & 4.89E-01 & 4.71E-01 &  10.3 & 2.92 & 3.05E-02 & 1.75         &  21 \\
16 & 1.35 & 0.623 & 3.90E-03 & 1.91E-02 & 1.00E-02 & 4.86E-01 & 3.90E-01 &  9.5  & 2.92 & 3.15E-02 & 1.89         &  21 \\
\cline{1-13} & \multicolumn{3}{r}{$M_{\rm TDU}^{\rm Tot}$ =
\textbf{7.82E-02 $M_\odot$}} & & & & \multicolumn{3}{r}{$\tau_C$ =
\textbf{9.27E+05 yr}}& & &
\\
\\
\cutinhead{\textbf{$M$=3.0 $M_\odot$ }}
-3 & 2.98 & 0.550 & 0.00E+00 & 2.96E-02 & 2.50E-03 & 9.54E-01 & 0.00E+00 &  9.7  & 2.06 & 2.02E-02 & 0.31          & \nodata  \\
-2$^{(*)}$ & 2.97 & 0.555 & 0.00E+00 & 3.04E-02 & 4.40E-03 & 8.81E-01 & 0.00E+00 &  13.4 & 2.24 & 2.02E-02 & 0.31  & \nodata  \\
-1 & 2.97 & 0.560 & 0.00E+00 & 2.96E-02 & 5.50E-03 & 8.33E-01 & 0.00E+00 &  15.3 & 2.35 & 2.02E-02 & 0.31          & \nodata  \\
1  & 2.96 & 0.567 & 8.00E-04 & 2.89E-02 & 6.80E-03 & 7.81E-01 & 1.18E-01 &  16.0 & 2.46 & 2.02E-02 & 0.32          & 2 \\
2  & 2.96 & 0.574 & 2.30E-03 & 2.77E-02 & 7.80E-03 & 7.34E-01 & 2.95E-01 &  15.9 & 2.55 & 2.04E-02 & 0.35          & 16 \\
3  & 2.94 & 0.581 & 3.60E-03 & 2.69E-02 & 9.10E-03 & 6.74E-01 & 3.96E-01 &  15.7 & 2.63 & 2.08E-02 & 0.41          & 35 \\
4  & 2.93 & 0.587 & 4.90E-03 & 2.60E-02 & 9.90E-03 & 6.31E-01 & 4.95E-01 &  15.4 & 2.71 & 2.14E-02 & 0.48          & 33 \\
5  & 2.91 & 0.593 & 5.90E-03 & 2.53E-02 & 1.08E-02 & 5.81E-01 & 5.46E-01 &  15.0 & 2.77 & 2.20E-02 & 0.57          & 31 \\
6  & 2.87 & 0.598 & 6.70E-03 & 2.45E-02 & 1.14E-02 & 5.45E-01 & 5.88E-01 &  14.5 & 2.82 & 2.26E-02 & 0.66          & 27 \\
7  & 2.81 & 0.603 & 7.00E-03 & 2.37E-02 & 1.18E-02 & 5.11E-01 & 5.93E-01 &  13.9 & 2.87 & 2.33E-02 & 0.76          & 26 \\
8  & 2.74 & 0.608 & 7.00E-03 & 2.30E-02 & 1.18E-02 & 4.91E-01 & 5.93E-01 &  13.3 & 2.91 & 2.40E-02 & 0.86          & 25 \\
9  & 2.63 & 0.613 & 7.10E-03 & 2.21E-02 & 1.17E-02 & 4.78E-01 & 6.07E-01 &  12.5 & 2.93 & 2.47E-02 & 0.96          & 25 \\
10 & 2.49 & 0.617 & 6.80E-03 & 2.13E-02 & 1.16E-02 & 4.61E-01 & 5.86E-01 &  11.7 & 2.95 & 2.54E-02 & 1.07          & 25 \\
11 & 2.32 & 0.622 & 6.50E-03 & 2.05E-02 & 1.13E-02 & 4.55E-01 & 5.75E-01 &  11.0 & 2.97 & 2.62E-02 & 1.18          & 24 \\
12 & 2.10 & 0.626 & 6.20E-03 & 1.97E-02 & 1.09E-02 & 4.52E-01 & 5.74E-01 &  10.3 & 2.99 & 2.71E-02 & 1.30          & 22 \\
13 & 1.84 & 0.631 & 5.70E-03 & 1.89E-02 & 1.07E-02 & 4.41E-01 & 5.33E-01 &  9.6  & 2.99 & 2.80E-02 & 1.43          & 22 \\
14 & 1.55 & 0.635 & 4.90E-03 & 1.80E-02 & 1.01E-02 & 4.44E-01 & 4.85E-01 &  8.8  & 3.00 & 2.91E-02 & 1.59          & 21 \\
\cline{1-13} & \multicolumn{3}{r}{$M_{\rm TDU}^{\rm Tot}$ =
\textbf{7.54E-02 $M_\odot$}} & & & & \multicolumn{3}{r}{$\tau_C$ =
\textbf{4.62E+05 yr}}& & &
\\
\\
\cline{1-13}\cline{1-13} \multicolumn{13}{l}{$^a$Pulse number
(negative values correspond to TP not followed by TDU)}\\
\multicolumn{13}{l}{ $^bM_\odot$} \\
\multicolumn{13}{l}{ $^c10^4$ yr}\\ \multicolumn{13}{l}{$^d10^8$
K}\\
\multicolumn{13}{l}{$^{(*)} M_{\rm H}$ of this TP corresponds
to $M_{\rm He}^{3}$ of Table \ref{tab2}}\\
\enddata \label{tab3}
\end{deluxetable}

\begin{deluxetable}{lcccccccccccc}
\rotate \tablecolumns{13} \tablewidth{0pc} \tablecaption{Evolution
of selected physical parameters of the models at $Z=6\times
10^{-3}$.} \tablehead{ \colhead{$n_{\rm TP}^a$} & \colhead{$M_{\rm
Tot}^{b}$} & \colhead{$M_{\rm H}^{b}$} & \colhead{$\Delta M_{\rm
TDU}^{b}$} & \colhead{$\Delta M_{\rm CZ}^{b}$} & \colhead{$\Delta
M_{\rm H}^{b}$} & \colhead{$r$} & \colhead{$\lambda$}
&\colhead{$\Delta t_{ip}^{c}$} & \colhead{$T_{MAX}^{d}$} &
\colhead{Z$_{surf}$} & \colhead{C/O$_{\rm surf}$} &
\colhead{$n_c$}}\startdata \cutinhead{\textbf{$M$=1.5 $M_\odot$}}
-3 & 1.36 & 0.552 & 0.00E+00 & 3.13E-02 & 2.50E-03 & 9.59E-01 & 0.00E+00 &  10.2 & 2.07 & 6.06E-03 & 0.32        & \nodata\\
-2$^{(*)}$  & 1.35 & 0.556 & 0.00E+00 & 3.18E-02 & 4.50E-03 & 8.81E-01 & 0.00E+00 &  15.6 & 2.26 & 6.06E-03& 0.32& \nodata\\
-1 & 1.34 & 0.563 & 0.00E+00 & 3.11E-02 & 6.80E-03 & 8.17E-01 & 0.00E+00 &  17.9 & 2.40 & 6.06E-03 & 0.32        & \nodata\\
1  & 1.33 & 0.570 & 2.00E-04 & 2.97E-02 & 6.90E-03 & 7.87E-01 & 2.90E-02 &  18.1 & 2.51 & 6.06E-03 & 0.32        & 42\\
2  & 1.31 & 0.578 & 1.10E-03 & 2.78E-02 & 8.30E-03 & 7.29E-01 & 1.33E-01 &  17.2 & 2.57 & 6.25E-03 & 0.42        & 120\\
3  & 1.29 & 0.585 & 1.60E-03 & 2.64E-02 & 8.70E-03 & 6.81E-01 & 1.84E-01 &  16.1 & 2.65 & 6.78E-03 & 0.66        & 130\\
4  & 1.26 & 0.593 & 2.20E-03 & 2.50E-02 & 9.20E-03 & 6.39E-01 & 2.39E-01 &  14.9 & 2.71 & 7.62E-03 & 1.05        & 120\\
5  & 1.24 & 0.600 & 2.70E-03 & 2.37E-02 & 9.60E-03 & 6.04E-01 & 2.81E-01 &  13.7 & 2.75 & 8.67E-03 & 1.53        & 110\\
6  & 1.20 & 0.607 & 2.80E-03 & 2.24E-02 & 9.60E-03 & 5.78E-01 & 2.92E-01 &  12.5 & 2.79 & 9.84E-03 & 2.05        & 97\\
7  & 1.16 & 0.614 & 2.80E-03 & 2.12E-02 & 9.50E-03 & 5.59E-01 & 2.95E-01 &  11.4 & 2.82 & 1.11E-02 & 2.59        & 91\\
8  & 1.09 & 0.620 & 2.40E-03 & 2.01E-02 & 9.20E-03 & 5.44E-01 & 2.62E-01 &  10.4 & 2.84 & 1.23E-02 & 3.16        & 87\\
\cline{1-13} & \multicolumn{3}{r}{$M_{\rm TDU}^{\rm Tot}$ =
\textbf{1.58E-02 $M_\odot$}} & & & & \multicolumn{3}{r}{$\tau_C$ =
\textbf{7.86E+05 yr}}& & &
\\
\\
\cutinhead{\textbf{$M$=2.0 $M_\odot$}}
-2$^{(*)}$ & 1.97 & 0.555 & 0.00E+00 & 3.32E-02 & 4.70E-03 & 8.82E-01 & 0.00E+00  & 1 5.9 & 2.27 & 6.07E-03 & 0.29& \nodata \\
-1 & 1.97 & 0.561 & 0.00E+00 & 3.20E-02 & 6.30E-03 & 8.21E-01 & 0.00E+00  & 2 0.3 & 2.42 & 6.07E-03 & 0.29        & \nodata \\
1  & 1.96 & 0.569 & 1.60E-03 & 3.06E-02 & 7.90E-03 & 7.58E-01 & 2.03E-01  & 2 0.9 & 2.55 & 6.29E-03 & 0.40        & 110 \\
2  & 1.95 & 0.577 & 3.40E-03 & 2.91E-02 & 9.50E-03 & 6.87E-01 & 3.58E-01  & 2 0.4 & 2.65 & 6.91E-03 & 0.69        & 140 \\
3  & 1.93 & 0.585 & 5.30E-03 & 2.80E-02 & 1.09E-02 & 6.25E-01 & 4.86E-01  & 1 9.6 & 2.74 & 7.91E-03 & 1.15        & 110 \\
4  & 1.91 & 0.591 & 6.70E-03 & 2.70E-02 & 1.20E-02 & 5.68E-01 & 5.58E-01  & 1 8.8 & 2.82 & 9.13E-03 & 1.70        & 110 \\
5  & 1.89 & 0.597 & 7.30E-03 & 2.60E-02 & 1.28E-02 & 5.19E-01 & 5.70E-01  & 17.7 & 2.87  & 1.04E-02 & 2.27        & 94 \\
6  & 1.86 & 0.603 & 7.60E-03 & 2.50E-02 & 1.29E-02 & 4.93E-01 & 5.89E-01  & 16.5 & 2.90 & 1.18E-02 & 2.85         & 91 \\
7  & 1.81 & 0.608 & 7.60E-03 & 2.40E-02 & 1.29E-02 & 4.71E-01 & 5.89E-01  & 15.3 & 2.93 & 1.31E-02 & 3.42         & 87 \\
8  & 1.75 & 0.613 & 7.40E-03 & 2.30E-02 & 1.27E-02 & 4.59E-01 & 5.83E-01  & 14.1 & 2.96 & 1.45E-02 & 3.99         & 85 \\
9  & 1.66 & 0.618 & 7.10E-03 & 2.20E-02 & 1.22E-02 & 4.58E-01 & 5.82E-01  & 12.9 & 2.99 & 1.58E-02 & 4.56         & 84 \\
10 & 1.55 & 0.623 & 6.50E-03 & 2.10E-02 & 1.18E-02 & 4.45E-01 & 5.51E-01  & 11.9 & 3.00 & 1.72E-02 & 5.13         & 82 \\
11 & 1.40 & 0.628 & 5.50E-03 & 1.98E-02 & 1.14E-02 & 4.34E-01 & 4.82E-01  & 10.8 & 3.00 & 1.87E-02 & 5.71         & 80 \\
12 & 1.21 & 0.633 & 4.50E-03 & 1.87E-02 & 1.06E-02 & 4.45E-01 & 4.25E-01  & 9.68 & 3.00 & 2.02E-02 & 6.35         & 79 \\
13 & 1.00 & 0.638 & 1.50E-03 & 1.75E-02 & 9.60E-03 & 4.60E-01 & 1.56E-01  & 8.47 & 2.99 & 2.11E-02 & 6.67         & 65 \\
14 & 0.83 & 0.644 & 7.00E-04 & 1.54E-02 & 7.80E-03 & 5.04E-01 & 9.00E-02  & 6.91 & 2.92 & 2.17E-02 & 6.91         & 63 \\
\cline{1-13} & \multicolumn{3}{r}{$M_{\rm TDU}^{\rm Tot}$ =
\textbf{7.27E-02 $M_\odot$}} & & & & \multicolumn{3}{r}{$\tau_C$ =
\textbf{1.38E+06 yr}}& & &
\\
\\
\cutinhead{\textbf{$M$=2.5 $M_\odot$}}
-1$^{(*)}$& 2.47 & 0.549 & 0.00E+00 & 3.44E-02 & 4.90E-03 & 8.78E-01 & 0.00E+00 & 18.1 & 2.28 &6.06E-03&0.29& \nodata  \\
1 & 2.47 & 0.555 & 3.00E-04 & 3.30E-02 & 6.70E-03 & 8.16E-01 & 4.48E-02 & 22.1 & 2.42 & 6.06E-03 & 0.29     & 31 \\
2 & 2.46 & 0.563 & 2.20E-03 & 3.16E-02 & 8.30E-03 & 7.52E-01 & 2.65E-01 & 22.4 & 2.54 & 6.32E-03 & 0.42     & 140 \\
3 & 2.45 & 0.571 & 4.50E-03 & 3.04E-02 & 9.90E-03 & 6.85E-01 & 4.55E-01 & 21.9 & 2.65 & 6.94E-03 & 0.72     & 140 \\
4 & 2.44 & 0.578 & 6.30E-03 & 2.93E-02 & 1.16E-02 & 6.17E-01 & 5.43E-01 & 21.3 & 2.74 & 7.81E-03 & 1.13     & 110 \\
5 & 2.42 & 0.585 & 7.60E-03 & 2.84E-02 & 1.27E-02 & 5.63E-01 & 5.98E-01 & 20.5 & 2.81 & 8.82E-03 & 1.60     & 95 \\
6 & 2.40 & 0.591 & 8.50E-03 & 2.75E-02 & 1.35E-02 & 5.20E-01 & 6.30E-01 & 19.4 & 2.87 & 9.88E-03 & 2.08     & 87 \\
7 & 2.38 & 0.596 & 8.80E-03 & 2.66E-02 & 1.39E-02 & 4.86E-01 & 6.33E-01 & 18.3 & 2.91 & 1.10E-02 & 2.56     & 75 \\
8 & 2.36 & 0.601 & 9.00E-03 & 2.56E-02 & 1.39E-02 & 4.67E-01 & 6.47E-01 & 17.1 & 2.95 & 1.20E-02 & 3.03     & 73 \\
9 & 2.33 & 0.606 & 9.10E-03 & 2.47E-02 & 1.38E-02 & 4.49E-01 & 6.59E-01 & 15.9 & 2.98 & 1.31E-02 & 3.50     & 73 \\
10& 2.29 & 0.610 & 8.90E-03 & 2.38E-02 & 1.37E-02 & 4.32E-01 & 6.50E-01 & 14.8 & 2.99 & 1.41E-02 & 3.97     & 72 \\
11& 2.23 & 0.615 & 8.70E-03 & 2.29E-02 & 1.34E-02 & 4.25E-01 & 6.49E-01 & 13.8 & 3.02 & 1.52E-02 & 4.43     & 71 \\
12& 2.16 & 0.619 & 8.50E-03 & 2.20E-02 & 1.30E-02 & 4.20E-01 & 6.54E-01 & 12.8 & 3.04 & 1.63E-02 & 4.88     & 69 \\
13& 2.07 & 0.623 & 8.10E-03 & 2.12E-02 & 1.27E-02 & 4.11E-01 & 6.38E-01 & 11.9 & 3.05 & 1.73E-02 & 5.34     & 66 \\
14& 1.95 & 0.628 & 7.70E-03 & 2.03E-02 & 1.22E-02 & 4.09E-01 & 6.31E-01 & 11.1 & 3.06 & 1.85E-02 & 5.81     & 79 \\
15& 1.80 & 0.632 & 7.10E-03 & 1.95E-02 & 1.18E-02 & 4.05E-01 & 6.02E-01 & 10.3 & 3.06 & 1.96E-02 & 6.29     & 75 \\
16& 1.62 & 0.636 & 6.30E-03 & 1.87E-02 & 1.12E-02 & 4.10E-01 & 5.63E-01 & 9.5  & 3.07 & 2.09E-02 & 6.80     & 72 \\
17& 1.39 & 0.640 & 5.10E-03 & 1.77E-02 & 1.05E-02 & 4.17E-01 & 4.86E-01 & 8.7  & 3.06 & 2.22E-02 & 7.35     & 83 \\
18& 1.13 & 0.644 & 3.40E-03 & 1.66E-02 & 9.60E-03 & 4.31E-01 & 3.54E-01 & 7.8  & 3.04 & 2.36E-02 & 7.90     & 70 \\
19& 0.89 & 0.649 & 1.10E-03 & 1.51E-02 & 8.30E-03 & 4.58E-01 & 1.33E-01 & 6.5  & 3.00 & 2.45E-02 & 8.25     & 70   \\
\cline{1-13} & \multicolumn{3}{r}{$M_{\rm TDU}^{\rm Tot}$ =
\textbf{1.21E-01 $M_\odot$}} & & & & \multicolumn{3}{r}{$\tau_C$ =
\textbf{1.99E+06 yr}}& & &
\\
\\
\cutinhead{\textbf{$M$=3.0 $M_\odot$}}
-2& 2.96 & 0.653 & 0.00E+00 & 1.67E-02 & \nodata  & \nodata  & \nodata  & \nodata & 2.23 & 6.06E-03 & 0.29 & \nodata\\
-1$^{(*)}$& 2.96 & 0.656 & 0.00E+00 & 1.66E-02 & 3.40E-03 & 8.24E-01 & 0.00E+00 & 5.2 & 2.41 & 6.06E-03 & 0.29 & \nodata\\
1 & 2.95 & 0.661 & 8.00E-04 & 1.63E-02 & 4.40E-03 & 7.47E-01 & 1.82E-01 & 5.9 & 2.57 & 6.11E-03 & 0.31     & 82  \\
2 & 2.94 & 0.665 & 2.20E-03 & 1.57E-02 & 5.40E-03 & 6.74E-01 & 4.07E-01 & 6.1 & 2.69 & 6.31E-03 & 0.42     & 100  \\
3 & 2.93 & 0.670 & 3.20E-03 & 1.54E-02 & 6.40E-03 & 5.96E-01 & 5.00E-01 & 6.2 & 2.79 & 6.66E-03 & 0.59     & 100 \\
4 & 2.89 & 0.673 & 4.10E-03 & 1.51E-02 & 7.00E-03 & 5.44E-01 & 5.86E-01 & 6.3 & 2.89 & 7.13E-03 & 0.83     & 100  \\
5 & 2.84 & 0.677 & 4.90E-03 & 1.49E-02 & 7.60E-03 & 4.94E-01 & 6.45E-01 & 6.4 & 2.97 & 7.69E-03 & 1.10     & 100  \\
6 & 2.77 & 0.680 & 5.60E-03 & 1.46E-02 & 8.20E-03 & 4.50E-01 & 6.83E-01 & 6.4 & 3.03 & 8.32E-03 & 1.41     & 99  \\
7 & 2.66 & 0.683 & 5.90E-03 & 1.44E-02 & 8.50E-03 & 4.14E-01 & 6.94E-01 & 6.3 & 3.09 & 9.01E-03 & 1.75     & 97  \\
8 & 2.52 & 0.686 & 6.20E-03 & 1.42E-02 & 8.70E-03 & 3.92E-01 & 7.13E-01 & 6.2 & 3.14 & 9.77E-03 & 2.11     & 94  \\
9 & 2.34 & 0.689 & 6.30E-03 & 1.39E-02 & 8.90E-03 & 3.70E-01 & 7.08E-01 & 6.0 & 3.16 & 1.06E-02 & 2.51     & 82  \\
10& 2.11 & 0.691 & 6.10E-03 & 1.36E-02 & 8.80E-03 & 3.59E-01 & 6.93E-01 & 5.8 & 3.20 & 1.16E-02 & 2.97     & 100  \\
11& 1.82 & 0.694 & 5.60E-03 & 1.33E-02 & 8.70E-03 & 3.53E-01 & 6.44E-01 & 5.6 & 3.21 & 1.27E-02 & 3.49     & 110  \\
12& 1.45 & 0.696 & 4.10E-03 & 1.28E-02 & 8.30E-03 & 3.58E-01 & 4.94E-01 & 5.3 & 3.22 & 1.39E-02 & 4.05     & 110  \\
13& 1.08 & 0.700 & 1.70E-03 & 1.18E-02 & 7.40E-03 & 3.79E-01 & 2.30E-01 & 4.7 & 3.20 & 1.47E-02 & 4.54     & 100  \\
\cline{1-13} & \multicolumn{3}{r}{$M_{\rm TDU}^{\rm Tot}$ =
\textbf{5.67E-02 $M_\odot$}} & & & & \multicolumn{3}{r}{$\tau_C$ =
\textbf{4.93E+05 yr}}& & &
\\
\\
\cline{1-13}\cline{1-13} \multicolumn{12}{l}{$^a$Pulse number
(negative values correspond to TP not followed by TDU)}\\
\multicolumn{13}{l}{ $^bM_\odot$} \\
\multicolumn{13}{l}{ $^c10^4$ yr}\\ \multicolumn{13}{l}{$^d10^8$
K}\\
\multicolumn{13}{l}{$^{(*)} M_{\rm H}$ of this TP corresponds
to $M_{\rm He}^{3}$ of Table \ref{tab2}}\\
\enddata \label{tab4}
\end{deluxetable}

\begin{deluxetable}{lcccccccccccc}
\rotate \tablecolumns{13} \tablewidth{0pc} \tablecaption{Evolution
of selected physical parameters of the models at $Z=1\times
10^{-3}$.} \tablehead{ \colhead{$n_{\rm TP}^a$} & \colhead{$M_{\rm
Tot}^{b}$} & \colhead{$M_{\rm H}^{b}$} & \colhead{$\Delta M_{\rm
TDU}^{b}$} & \colhead{$\Delta M_{\rm CZ}^{b}$} & \colhead{$\Delta
M_{\rm H}^{b}$} & \colhead{$r$} & \colhead{$\lambda$}
&\colhead{$\Delta t_{ip}^{c}$} & \colhead{$T_{MAX}^{d}$} &
\colhead{Z$_{surf}$} & \colhead{C/O$_{\rm surf}$} &
\colhead{$n_c$}}\startdata \cutinhead{\textbf{$M$=1.5 $M_\odot$}}
-2 & 1.40 & 0.566 & 0.00E+00 & 2.81E-02 & 1.70E-03 & 1.08E+00 & 0.00E+00 &  7.8  & 2.00 & 1.01E-03 & 0.29 & \nodata \\
-1$^{(*)}$ & 1.39 & 0.572 & 0.00E+00 & 3.16E-02 & 5.50E-03 & 8.47E-01 & 0.00E+00 &  17.6 & 2.40 & 1.01E-03 & 0.29 & \nodata  \\
1  & 1.39 & 0.579 & 2.30E-03 & 3.01E-02 & 7.50E-03 & 7.68E-01 & 3.07E-01 &  22.2 & 2.58 & 1.58E-03 & 1.95      & 740\\
2  & 1.38 & 0.587 & 4.50E-03 & 2.84E-02 & 9.70E-03 & 6.73E-01 & 4.64E-01 &  21.1 & 2.70 & 3.01E-03 & 5.51      & 830\\
3  & 1.36 & 0.593 & 5.90E-03 & 2.71E-02 & 1.12E-02 & 5.99E-01 & 5.27E-01 &  19.3 & 2.79 & 4.96E-03 & 9.34      & 710\\
4  & 1.34 & 0.599 & 6.80E-03 & 2.59E-02 & 1.21E-02 & 5.44E-01 & 5.62E-01 &  17.7 & 2.85 & 7.19E-03 & 12.8      & 680\\
5  & 1.31 & 0.605 & 7.00E-03 & 2.48E-02 & 1.26E-02 & 5.03E-01 & 5.63E-01 &  16.2 & 2.90 & 9.33E-03 & 15.5      & 620\\
6  & 1.29 & 0.611 & 7.10E-03 & 2.36E-02 & 1.25E-02 & 4.86E-01 & 5.60E-01 &  14.6 & 2.93 & 1.19E-02 & 18.3      & 610\\
7  & 1.25 & 0.616 & 6.50E-03 & 2.24E-02 & 1.21E-02 & 4.68E-01 & 5.37E-01 &  13.2 & 2.96 & 1.41E-02 & 20.6      & 590\\
8  & 1.22 & 0.621 & 6.10E-03 & 2.13E-02 & 1.16E-02 & 4.64E-01 & 5.26E-01 &  11.9 & 2.97 & 1.63E-02 & 22.5      & 560\\
9  & 1.19 & 0.626 & 5.30E-03 & 2.01E-02 & 1.11E-02 & 4.56E-01 & 4.77E-01 &  10.7 & 2.98 & 1.84E-02 & 24.2      & 510\\
10 & 1.15 & 0.631 & 4.50E-03 & 1.90E-02 & 1.04E-02 & 4.61E-01 & 4.33E-01 &  9.6  & 2.98 & 2.03E-02 & 25.7      & 530\\
11 & 1.11 & 0.636 & 3.60E-03 & 1.78E-02 & 9.80E-03 & 4.62E-01 & 3.67E-01 &  8.6  & 2.97 & 2.20E-02 & 26.9      & 560\\
12 & 1.06 & 0.642 & 2.70E-03 & 1.68E-02 & 9.00E-03 & 4.73E-01 & 3.00E-01 &  7.7  & 2.96 & 2.35E-02 & 28.0      & 520\\
13 & 1.00 & 0.647 & 1.80E-03 & 1.57E-02 & 8.30E-03 & 4.79E-01 & 2.17E-01 &  6.9  & 2.95 & 2.47E-02 & 28.8      & 540\\
14 & 0.94 & 0.653 & 2.00E-04 & 1.47E-02 & 7.60E-03 & 4.92E-01 & 2.63E-02 &  6.1  & 2.93 & 2.50E-02 & 29.0      & 480 \\
\cline{1-13} & \multicolumn{3}{r}{$M_{\rm TDU}^{\rm Tot}$ =
\textbf{6.43E-02 $M_\odot$}} & & & & \multicolumn{3}{r}{$\tau_C$ =
\textbf{1.50E+06 yr}}& & &
\\
\\
\cutinhead{\textbf{$M$=2.0 $M_\odot$}}
-1$^{(*)}$ & 1.94 & 0.600 & 0.00E+00 & 2.36E-02 & 2.20E-03 & 9.55E-01 & 0.00E+00 & 6.48  & 2.13 & 1.01E-03 & 0.25 & \nodata\\
1  & 1.93 & 0.604 & 2.00E-04 & 2.53E-02 & 4.70E-03 & 8.23E-01 & 4.30E-02 & 12.6  & 2.47 & 1.03E-03 & 0.31      & 370\\
2  & 1.92 & 0.611 & 3.00E-03 & 2.41E-02 & 6.70E-03 & 7.31E-01 & 4.48E-01 & 15.1  & 2.64 & 1.56E-03 & 1.85      & 1000\\
3  & 1.91 & 0.617 & 5.40E-03 & 2.33E-02 & 8.80E-03 & 6.27E-01 & 6.14E-01 & 15.0  & 2.79 & 2.63E-03 & 4.56      & 730\\
4  & 1.89 & 0.622 & 7.20E-03 & 2.28E-02 & 1.06E-02 & 5.36E-01 & 6.79E-01 & 14.7  & 2.90 & 4.05E-03 & 7.64      & 670\\
5  & 1.88 & 0.627 & 8.50E-03 & 2.23E-02 & 1.19E-02 & 4.72E-01 & 7.14E-01 & 14.2  & 2.98 & 5.64E-03 & 10.6      & 650\\
6  & 1.85 & 0.631 & 9.30E-03 & 2.19E-02 & 1.26E-02 & 4.27E-01 & 7.38E-01 & 13.6  & 3.03 & 7.31E-03 & 13.3      & 670\\
7  & 1.83 & 0.635 & 9.70E-03 & 2.14E-02 & 1.31E-02 & 3.94E-01 & 7.40E-01 & 13.0  & 3.08 & 9.01E-03 & 15.9      & 650\\
8  & 1.81 & 0.638 & 9.70E-03 & 2.09E-02 & 1.31E-02 & 3.78E-01 & 7.40E-01 & 12.2  & 3.11 & 1.07E-02 & 18.3      & 630\\
9  & 1.78 & 0.641 & 9.70E-03 & 2.03E-02 & 1.29E-02 & 3.70E-01 & 7.52E-01 & 11.5  & 3.15 & 1.24E-02 & 20.5      & 630\\
10 & 1.75 & 0.644 & 9.50E-03 & 1.97E-02 & 1.27E-02 & 3.58E-01 & 7.48E-01 & 10.8  & 3.16 & 1.41E-02 & 22.5      & 670\\
11 & 1.71 & 0.647 & 9.20E-03 & 1.91E-02 & 1.25E-02 & 3.51E-01 & 7.36E-01 & 10.1  & 3.18 & 1.58E-02 & 24.4      & 640\\
12 & 1.66 & 0.650 & 8.80E-03 & 1.85E-02 & 1.20E-02 & 3.51E-01 & 7.33E-01 & 9.46  & 3.18 & 1.74E-02 & 26.2      & 660\\
13 & 1.61 & 0.653 & 8.40E-03 & 1.79E-02 & 1.17E-02 & 3.48E-01 & 7.18E-01 & 8.83  & 3.18 & 1.91E-02 & 27.8      & 650\\
14 & 1.53 & 0.656 & 8.00E-03 & 1.72E-02 & 1.13E-02 & 3.50E-01 & 7.08E-01 & 8.24  & 3.19 & 2.08E-02 & 29.4      & 620\\
15 & 1.45 & 0.659 & 7.30E-03 & 1.66E-02 & 1.08E-02 & 3.50E-01 & 6.76E-01 & 7.65  & 3.18 & 2.26E-02 & 30.9      & 640\\
16 & 1.34 & 0.661 & 6.40E-03 & 1.58E-02 & 1.02E-02 & 3.56E-01 & 6.27E-01 & 7.06  & 3.17 & 2.44E-02 & 32.3      & 610\\
17 & 1.22 & 0.665 & 5.30E-03 & 1.50E-02 & 9.50E-03 & 3.68E-01 & 5.58E-01 & 6.46  & 3.17 & 2.62E-02 & 33.7      & 600\\
18 & 1.09 & 0.668 & 3.70E-03 & 1.41E-02 & 8.70E-03 & 3.82E-01 & 4.25E-01 & 5.83  & 3.13 & 2.80E-02 & 34.9      & 600\\
19 & 0.96 & 0.672 & 2.00E-03 & 1.31E-02 & 7.50E-03 & 4.11E-01 & 2.67E-01 & 5.14  & 3.10 & 2.96E-02 & 36.0      & 610\\
\cline{1-13} & \multicolumn{3}{r}{$M_{\rm TDU}^{\rm Tot}$ =
\textbf{1.31E-01 $M_\odot$}} & & & & \multicolumn{3}{r}{$\tau_C$ =
\textbf{1.85E+06 yr}}& & &
\\
\\
\cutinhead{\textbf{$M$=2.5 $M_\odot$}}
1  & 2.45 & 0.672 & 5.00E-04 & 1.56E-02 & 3.10E-03 & 8.30E-01 & 1.61E-01 &  4.8  & 2.45 & 1.02E-03 & 0.26      & 340 \\
2  & 2.45 & 0.676 & 2.30E-03 & 1.51E-02 & 4.40E-03 & 7.27E-01 & 5.23E-01 &  6.5  & 2.65 & 1.25E-03 & 0.96      & 440\\
3  & 2.44 & 0.680 & 4.20E-03 & 1.49E-02 & 6.00E-03 & 6.14E-01 & 7.00E-01 &  7.0  & 2.82 & 1.81E-03 & 2.52      & 480\\
4  & 2.43 & 0.683 & 5.70E-03 & 1.49E-02 & 7.40E-03 & 5.18E-01 & 7.70E-01 &  7.4  & 2.96 & 2.61E-03 & 4.58      & 600\\
5  & 2.42 & 0.686 & 6.80E-03 & 1.49E-02 & 8.50E-03 & 4.44E-01 & 8.00E-01 &  7.7  & 3.07 & 3.56E-03 & 6.78      & 780\\
6  & 2.41 & 0.688 & 7.70E-03 & 1.50E-02 & 9.30E-03 & 3.90E-01 & 8.28E-01 &  7.8  & 3.16 & 4.59E-03 & 8.98      & 820\\
7  & 2.38 & 0.690 & 8.20E-03 & 1.50E-02 & 9.90E-03 & 3.44E-01 & 8.28E-01 &  7.8  & 3.24 & 5.67E-03 & 11.1      & 900\\
8  & 2.35 & 0.693 & 8.60E-03 & 1.49E-02 & 1.03E-02 & 3.16E-01 & 8.35E-01 &  7.7  & 3.30 & 6.80E-03 & 13.3      & 910\\
9  & 2.30 & 0.695 & 9.00E-03 & 1.49E-02 & 1.06E-02 & 2.96E-01 & 8.49E-01 &  7.6  & 3.33 & 7.97E-03 & 15.3      & 940\\
10 & 2.24 & 0.696 & 9.10E-03 & 1.48E-02 & 1.08E-02 & 2.76E-01 & 8.43E-01 &  7.4  & 3.37 & 9.19E-03 & 17.4      & 1000\\
11 & 2.16 & 0.698 & 9.00E-03 & 1.46E-02 & 1.08E-02 & 2.67E-01 & 8.33E-01 &  7.2  & 3.40 & 1.05E-02 & 19.3      & 1100\\
12 & 2.07 & 0.700 & 9.10E-03 & 1.45E-02 & 1.07E-02 & 2.67E-01 & 8.50E-01 &  7.0  & 3.41 & 1.18E-02 & 21.3      & 1100\\
13 & 1.95 & 0.701 & 8.90E-03 & 1.42E-02 & 1.07E-02 & 2.55E-01 & 8.32E-01 &  6.7  & 3.43 & 1.32E-02 & 23.2      & 1100\\
14 & 1.81 & 0.703 & 8.60E-03 & 1.40E-02 & 1.05E-02 & 2.55E-01 & 8.19E-01 &  6.5  & 3.45 & 1.47E-02 & 25.1      & 1100\\
15 & 1.64 & 0.705 & 8.30E-03 & 1.37E-02 & 1.02E-02 & 2.59E-01 & 8.14E-01 &  6.2  & 3.45 & 1.64E-02 & 27.2      & 1100 \\
16 & 1.43 & 0.706 & 7.50E-03 & 1.33E-02 & 9.90E-03 & 2.56E-01 & 7.58E-01 &  5.8  & 3.44 & 1.83E-02 & 29.3      & 1100\\
17 & 1.18 & 0.708 & 5.80E-03 & 1.26E-02 & 9.30E-03 & 2.70E-01 & 6.24E-01 &  5.3  & 3.42 & 2.07E-02 & 31.8      & 1100\\
18 & 0.94 & 0.710 & 2.40E-03 & 1.14E-02 & 7.90E-03 & 3.10E-01 & 3.04E-01 &  4.5  & 3.36 & 2.29E-02 & 33.8      & 970\\
\cline{1-13} & \multicolumn{3}{r}{$M_{\rm TDU}^{\rm Tot}$ =
\textbf{1.22E-01 $M_\odot$}} & & & & \multicolumn{3}{r}{$\tau_C$ =
\textbf{1.05E+06 yr}}& & &
\\
\\
\cutinhead{\textbf{$M$=3.0 $M_\odot$}}
-1$^{(*)}$ & 2.95 & 0.777 & 0.00E+00 & 7.35E-03 & \nodata & \nodata & \nodata    &  \nodata & 2.35 & 1.01E-03 & 0.27 & \nodata\\
1  & 2.95 & 0.779 & 9.00E-04 & 7.49E-03 & 1.90E-03 & 7.85E-01 & 4.74E-01 &  1.8  & 2.55 & 1.06E-03 & 0.43     & 460\\
2  & 2.94 & 0.780 & 1.90E-03 & 7.25E-03 & 2.50E-03 & 6.77E-01 & 7.60E-01 &  1.9  & 2.71 & 1.25E-03 & 0.97     & 740\\
3  & 2.94 & 0.782 & 2.70E-03 & 7.14E-03 & 3.10E-03 & 5.66E-01 & 8.71E-01 &  2.1  & 2.85 & 1.60E-03 & 1.86     & 720\\
4  & 2.92 & 0.783 & 3.30E-03 & 7.18E-03 & 3.80E-03 & 4.87E-01 & 8.46E-01 &  2.3  & 2.99 & 2.00E-03 & 2.86     & 840\\
5  & 2.89 & 0.784 & 3.60E-03 & 7.17E-03 & 4.30E-03 & 4.18E-01 & 8.37E-01 &  2.4  & 3.10 & 2.43E-03 & 3.90     & 920\\
6  & 2.85 & 0.785 & 3.80E-03 & 7.18E-03 & 4.60E-03 & 3.73E-01 & 8.26E-01 &  2.5  & 3.18 & 2.89E-03 & 4.99     & 1000\\
7  & 2.79 & 0.786 & 3.90E-03 & 7.24E-03 & 4.80E-03 & 3.39E-01 & 8.13E-01 &  2.5  & 3.26 & 3.39E-03 & 6.13     & 1200\\
8  & 2.72 & 0.787 & 4.00E-03 & 7.29E-03 & 5.00E-03 & 3.19E-01 & 8.00E-01 &  2.6  & 3.32 & 3.94E-03 & 7.32     & 1300\\
9  & 2.62 & 0.788 & 4.30E-03 & 7.35E-03 & 5.10E-03 & 3.17E-01 & 8.43E-01 &  2.6  & 3.38 & 4.53E-03 & 8.57     & 1600\\
10 & 2.51 & 0.789 & 4.40E-03 & 7.37E-03 & 5.30E-03 & 2.88E-01 & 8.30E-01 &  2.6  & 3.43 & 5.18E-03 & 9.88     & 1900\\
11 & 2.38 & 0.790 & 4.50E-03 & 7.38E-03 & 5.40E-03 & 2.73E-01 & 8.33E-01 &  2.6  & 3.46 & 5.90E-03 & 11.3     & 2300\\
12 & 2.23 & 0.791 & 4.70E-03 & 7.39E-03 & 5.50E-03 & 2.61E-01 & 8.55E-01 &  2.6  & 3.49 & 6.73E-03 & 12.7     & 2900\\
13 & 2.03 & 0.792 & 4.80E-03 & 7.40E-03 & 5.70E-03 & 2.37E-01 & 8.42E-01 &  2.6  & 3.52 & 7.70E-03 & 14.3     & 3600\\
14 & 1.79 & 0.793 & 4.70E-03 & 7.41E-03 & 5.70E-03 & 2.32E-01 & 8.25E-01 &  2.5  & 3.56 & 8.90E-03 & 16.1     & 5000\\
15 & 1.51 & 0.794 & 3.90E-03 & 7.34E-03 & 5.60E-03 & 2.36E-01 & 6.96E-01 &  2.5  & 3.58 & 1.04E-02 & 18.1     & 5500\\
16 & 1.21 & 0.795 & 1.90E-03 & 6.95E-03 & 5.10E-03 & 2.78E-01 & 3.73E-01 &  2.3  & 3.54 & 1.13E-02 & 19.7     & 4700\\
\cline{1-13} & \multicolumn{3}{r}{$M_{\rm TDU}^{\rm Tot}$ =
\textbf{5.73E-02 $M_\odot$}} & & & & \multicolumn{3}{r}{$\tau_C$ =
\textbf{1.44E+05 yr}}& & &
\\
\\
\cline{1-13}\cline{1-13} \multicolumn{13}{l}{$^a$Pulse number
(negative values correspond to TP not followed by TDU)}\\
\multicolumn{13}{l}{ $^bM_\odot$} \\
\multicolumn{13}{l}{ $^c10^4$ yr}\\ \multicolumn{13}{l}{$^d10^8$
K}\\
\multicolumn{13}{l}{$^{(*)} M_{\rm H}$ of this TP corresponds
to $M_{\rm He}^{3}$ of Table \ref{tab2}}\\
\enddata \label{tab5}
\end{deluxetable}

\begin{deluxetable}{lcccccccccccc}
\rotate \tablecolumns{13} \tablewidth{0pc} \tablecaption{Final
enrichments of the three {\it s}-process peaks ([ls/Fe], [hs/Fe]
and [Pb/Fe]) and {\it s}-process indexes ([hs/ls] and [Pb/hs]) for
different masses at selected metallicities. The final surface
composition of some key light elements, such as the final surface
C/O, C/N and $^{12}$C/$^{13}$C ratios are given.}
\tablehead{\colhead{} & \multicolumn{4}{c}{$Z=2.0\times 10^{-2}$}&
\multicolumn{4}{c}{$Z=6.0\times 10^{-3}$}&
\multicolumn{4}{c}{$Z=10^{-3}$}} \startdata
$M/M_\odot$ & 1.5 & 2.0 & 2.5 & 3.0 & 1.5 & 2.0 & 2.5 & 3.0 & 1.5 & 2.0 & 2.5 & 3.0 \\
\cline{1-13}
$\rm[ls/Fe]$       & 0.40 & 0.93  & 1.12  &  1.02  &     0.89  & 1.36  & 1.44  &  1.03  & 1.26  & 1.35  & 1.21  &  1.03  \\
$\rm[hs/Fe]$       & 0.10 & 0.45  & 0.58  &  0.51  &     1.40  & 1.75  & 1.79  &  1.56  & 1.91  & 1.98  & 1.72  &  1.37  \\
$\rm[Pb/Fe]$       & 0.02 & 0.19  & 0.27  &  0.24  &     1.30  & 1.59  & 1.61  &  1.46  & 2.86  & 2.87  & 2.75  &  2.62  \\
$\rm[hs/ls]$       & -0.30 & -0.48  & -0.54  &  -0.51  & 0.50  & 0.39  & 0.34  & 0.53  & 0.65  & 0.63  & 0.52  &  0.35  \\
$\rm[Pb/hs]$       & -0.09 & -0.26  & -0.31  &  -0.27  & -0.10 & -0.16  & -0.18  &  -0.10 & 0.95 & 0.89 & 1.03 & 1.25 \\
\cline{1-13}
$\rm[C/Fe] $       & 0.16 & 0.42  & 0.56  & 0.48  & 0.82  & 1.17   & 1.23  &  0.94  & 2.11  & 2.18  & 2.08  &  1.78  \\
$\rm[N/Fe] $       & 0.32 & 0.35  & 0.39  &  0.42  & 0.36  & 0.40  & 0.44  &  0.47  & 0.63  & 0.56  & 0.50  &  0.49  \\
$\rm[O/Fe] $       & 0.00 & 0.00  & -0.02  &  -0.03 & 0.02  & 0.03 & 0.01  &  -0.01 & 0.35  & 0.32  & 0.25  &  0.18  \\
$\rm[F/Fe] $       & 0.12 & 0.40  & 0.59  &  0.50  & 0.55  & 1.06  & 1.18  &  0.81  & 2.08  & 2.22  & 1.99  &  1.51  \\
$\rm[Ne/Fe]$       & 0.05 & 0.17  & 0.27  &  0.22  & 0.20  & 0.56  & 0.67  &  0.32  & 1.42  & 1.60  & 1.38  &  0.80  \\
$\rm[Na/Fe]$       & 0.05 & 0.16  & 0.28  &  0.21  & 0.11  & 0.36  & 0.46  &  0.23  & 1.13  & 1.28  & 1.01  &  0.61  \\
$\rm[Mg/Fe]$       & 0.00 & 0.00  & 0.01  &  0.01  & 0.01  & 0.07  & 0.11  &  0.05  & 0.43  & 0.72  & 0.73  &  0.42  \\
\cline{1-13}
(C/O)$_{\rm surf}$ &              0.71   & 1.33  & 1.89  &  1.59  & 3.15  & 6.92  & 8.25  &  4.55  & 29.0  & 35.9  & 33.9  &  19.7  \\
(C/N)$_{\rm surf}$ &              2.48   & 4.22  & 5.34  &  4.08  & 10.5  & 21.3  & 22.1  &  10.9  & 111  & 151   & 157   &  70.7  \\
($^{12}$C/$^{13}$C)$_{\rm surf}$  & 50   &  98   & 141   &  115   &  239  & 575   & 693   &  349   & 3600  & 5193  & 4302  &  196   \\
\enddata
\label{tab6}
\end{deluxetable}

\begin{deluxetable}{lcccc}
\tablecolumns{5} \tablewidth{0pc} \tablecaption{Yields (in
$M_\odot$) of selected isotopes for $Z=2.0\times 10^{-2}$.}
\tablehead{\colhead{Isotope}  & \colhead{$M$=1.5 $M_\odot$} &
\colhead{$M$=2.0 $M_\odot$} & \colhead{$M$=2.5 $M_\odot$} &
\colhead{$M$=3.0 $M_\odot$}} \startdata \cline{1-5}
$^{1}$H     &  -2.25E-02   &   -6.08E-02   & -7.76E-02&   -9.01E-02   \\
$^{4}$He    &  2.04E-02    &   4.79E-02    & 6.06E-02 &   7.36E-02    \\
$^{12}$C    &  4.61E-04    &   8.22E-03    & 1.20E-02 &   1.10E-02    \\
$^{13}$C    &  5.38E-05    &   1.12E-04    & 9.81E-05 &   1.27E-04    \\
$^{14}$N    &  9.86E-04    &   2.66E-03    & 2.91E-03 &   4.16E-03    \\
$^{15}$N    &  -1.40E-06   &   -4.19E-06   & -4.18E-06&   -5.53E-06   \\
$^{16}$O    &  -1.88E-05   &   -1.51E-04   & -6.59E-04&   -1.21E-03   \\
$^{17}$O    &  4.97E-06    &   5.17E-05    & 7.03E-05 &   7.12E-05    \\
$^{18}$O    &  -3.64E-06   &   -1.14E-05   & -1.17E-05&   -1.54E-05   \\
$^{19}$F    &  1.03E-07    &   1.45E-06    & 2.40E-06 &   2.25E-06    \\
$^{20}$Ne   &  -8.64E-07   &   -4.24E-06   & -1.70E-06&   -2.85E-06   \\
$^{21}$Ne   &  2.58E-09    &   8.55E-08    & 3.61E-07 &   7.49E-07    \\
$^{22}$Ne   &  1.21E-04    &   1.34E-03    & 2.21E-03 &   2.03E-03    \\
$^{23}$Na   &  4.58E-06    &   3.95E-05    & 6.94E-05 &   6.67E-05    \\
$^{24}$Mg   &  9.25E-07    &   1.43E-05    & 2.96E-05 &   3.03E-05    \\
$^{25}$Mg   &  -4.76E-07   &   -3.63E-07   & 5.18E-07 &   3.56E-06    \\
$^{26}$Mg   &  4.66E-07    &   5.91E-06    & 1.05E-05 &   1.40E-05    \\
$^{26}$Al   &  1.76E-07    &   7.92E-07    & 7.87E-07 &   7.27E-07    \\
$^{36}$Cl   &  2.96E-10    &   3.07E-09    & 3.28E-09 &   4.78E-09    \\
$^{41}$Ca   &  5.57E-10    &   9.09E-09    & 9.70E-09 &   1.39E-08    \\
$^{60}$Fe   &  6.74E-09    &   5.26E-08    & 8.23E-08 &   4.33E-08    \\
$^{89}$Y    &  1.29E-08    &   1.90E-07    & 2.84E-07 &   2.81E-07    \\
$^{107}$Pd  &  1.17E-10    &   1.84E-09    & 2.39E-09 &   2.46E-09    \\
$^{139}$La  &  4.67E-10    &   1.01E-08    & 1.46E-08 &   1.45E-08    \\
$^{205}$Pb  &  3.75E-12    &   9.44E-11    & 1.39E-10 &   1.40E-10    \\
$^{208}$Pb  &  1.42E-10    &   7.23E-09    & 1.07E-08 &   1.21E-08    \\
\enddata
\label{tab7}
\end{deluxetable}

\begin{deluxetable}{lcccc}
\tablecolumns{5} \tablewidth{0pc} \tablecaption{Yields (in
$M_\odot$) of selected isotopes for $Z=6.0\times 10^{-3}$.}
\tablehead{\colhead{Isotope}  & \colhead{$M$=1.5 $M_\odot$} &
\colhead{$M$=2.0 $M_\odot$} & \colhead{$M$=2.5 $M_\odot$} &
\colhead{$M$=3.0 $M_\odot$}} \startdata \cline{1-5}
$^{1}$H     &   -2.79E-02   &   -6.59E-02   & -1.10E-01 &   -7.06E-02  \\
$^{4}$He    &   2.41E-02    &   5.09E-02    & 8.50E-02 &   5.77E-02   \\
$^{12}$C    &   2.85E-03    &   1.26E-02    & 2.09E-02 &   1.07E-02   \\
$^{13}$C    &   1.49E-05    &   2.01E-05    & 2.64E-05 &   3.46E-05   \\
$^{14}$N    &   3.34E-04    &   5.98E-04    & 1.03E-03 &   1.41E-03   \\
$^{15}$N    &   -4.55E-07   &   -9.14E-07   & -1.37E-06 &   -1.71E-06  \\
$^{16}$O    &   6.01E-05    &   1.69E-04    & 6.84E-05 &   -2.35E-04  \\
$^{17}$O    &   3.61E-06    &   2.31E-05    & 2.58E-05 &   2.62E-05   \\
$^{18}$O    &   -1.32E-06   &   -2.75E-06   & -4.21E-06 &   -5.05E-06  \\
$^{19}$F    &   2.39E-07    &   1.59E-06    & 3.05E-06 &   1.43E-06   \\
$^{20}$Ne   &   -1.16E-06   &   1.28E-06    & 8.31E-06 &   -3.67E-06  \\
$^{21}$Ne   &   2.96E-08    &   2.54E-07    & 7.32E-07 &   5.70E-07   \\
$^{22}$Ne   &   1.67E-04    &   1.20E-03    & 2.33E-03 &   7.65E-04   \\
$^{23}$Na   &   3.06E-06    &   1.75E-05    & 3.60E-05 &   2.02E-05   \\
$^{24}$Mg   &   2.99E-06    &   2.37E-05    & 5.32E-05 &   9.17E-06   \\
$^{25}$Mg   &   1.85E-07    &   6.12E-06    & 1.58E-05 &   1.28E-05   \\
$^{26}$Mg   &   5.28E-07    &   7.02E-06    & 1.75E-05 &   1.79E-05   \\
$^{26}$Al   &   8.57E-08    &   1.23E-07    & 1.35E-07 &   2.31E-07   \\
$^{36}$Cl   &   2.59E-10    &   1.41E-09    & 2.17E-09 &   2.59E-09   \\
$^{41}$Ca   &   6.65E-10    &   3.36E-09    & 4.96E-09 &   6.24E-09   \\
$^{60}$Fe   &   1.69E-08    &   5.00E-08    & 2.28E-07 &   4.31E-08   \\
$^{89}$Y    &   1.67E-08    &   1.00E-07    & 1.74E-07 &   7.13E-08   \\
$^{107}$Pd  &   3.16E-10    &   1.68E-09    & 2.88E-09 &   1.26E-09   \\
$^{139}$La  &   1.17E-08    &   5.38E-08    & 8.62E-08 &   5.11E-08   \\
$^{205}$Pb  &   2.96E-10    &   1.04E-09    & 1.45E-09 &   7.54E-10   \\
$^{208}$Pb  &   3.37E-08    &   1.50E-07    & 2.33E-07 &   1.69E-07   \\
\enddata
\label{tab8}
\end{deluxetable}

\begin{deluxetable}{lcccc}
\tablecolumns{5} \tablewidth{0pc} \tablecaption{Yields (in
$M_\odot$) of selected isotopes for $Z=10^{-3}$.}
\tablehead{\colhead{Isotope}  & \colhead{$M$=1.5 $M_\odot$} &
\colhead{$M$=2.0 $M_\odot$} & \colhead{$M$=2.5 $M_\odot$} &
\colhead{$M$=3.0 $M_\odot$}} \startdata \cline{1-5}
$^{1}$H     &    -6.11E-02  &   -1.15E-01   & -1.05E-01 &   -6.03E-02\\
$^{4}$He    &     4.82E-02  &   8.85E-02    & 8.00E-02 &   4.57E-02\\
$^{12}$C    &     1.11E-02  &   2.23E-02    & 2.20E-02 &   1.30E-02\\
$^{13}$C    &     2.86E-06  &   3.63E-06    & 4.75E-06 &   9.73E-05\\
$^{14}$N    &     9.81E-05  &   1.52E-04    & 1.83E-04 &   2.31E-04\\
$^{15}$N    &    -7.51E-08  &   -1.65E-07   & -2.44E-07 &   -2.84E-07\\
$^{16}$O    &     2.91E-04  &   4.47E-04    & 3.89E-04 &   3.05E-04\\
$^{17}$O    &     1.71E-06  &   7.09E-06    & 1.07E-05 &   5.29E-06\\
$^{18}$O    &    -2.67E-07  &   -5.07E-07   & -7.67E-07 &   -8.26E-07\\
$^{19}$F    &     1.71E-06  &   3.97E-06    & 2.97E-06 &   1.60E-06\\
$^{20}$Ne   &     6.35E-06  &   2.80E-05    & 1.57E-05 &   3.05E-06\\
$^{21}$Ne   &     2.45E-07  &   1.00E-06    & 1.22E-06 &   4.36E-07\\
$^{22}$Ne   &     1.05E-03  &   2.69E-03    & 1.87E-03 &   5.91E-04\\
$^{23}$Na   &     1.33E-05  &   3.25E-05    & 2.24E-05 &   1.03E-05\\
$^{24}$Mg   &     2.19E-05  &   7.33E-05    & 2.73E-05 &   3.35E-06\\
$^{25}$Mg   &     6.95E-06  &   4.36E-05    & 6.29E-05 &   3.21E-05\\
$^{26}$Mg   &     6.62E-06  &   3.14E-05    & 9.37E-05 &   4.81E-05\\
$^{26}$Al   &     2.57E-08  &   5.40E-08    & 6.74E-08 &   2.86E-08\\
$^{36}$Cl   &     1.91E-10  &   4.34E-10    & 5.05E-10 &   4.54E-10\\
$^{41}$Ca   &     3.77E-10  &   8.58E-10    & 9.62E-10 &   6.83E-10\\
$^{60}$Fe   &     2.33E-08  &   1.27E-07    & 5.40E-07 &   5.93E-07\\
$^{89}$Y    &     7.08E-09  &   1.55E-08    & 1.48E-08 &   1.00E-08\\
$^{107}$Pd  &     1.11E-10  &   1.99E-10    & 1.66E-10 &   1.25E-10\\
$^{139}$La  &     5.50E-09  &   1.08E-08    & 7.89E-09 &   4.39E-09\\
$^{205}$Pb  &     2.36E-10  &   3.95E-10    & 1.98E-10 &   1.42E-10\\
$^{208}$Pb  &     4.49E-07  &   8.48E-07    & 8.66E-07 &   7.95E-07\\
\enddata
\label{tab9}
\end{deluxetable}

\begin{deluxetable}{lccccccccccccc}
\rotate \tablecolumns{14} \tablewidth{0pc} \tablecaption{Surface
{\it s}-process branched number isotopic ratios for different
initial masses at selected metallicities. The tabulated ratios
relate to the branchings occurring at $^{85}$Kr, $^{86}$Rb,
$^{95}$Zr, $^{133}$Xe and $^{141}$Ce, respectively. Solar ratios
are also reported.} \tablehead{\colhead{Isot. Ratio} &
\colhead{$\odot$} & \multicolumn{4}{c}{$Z=2.0\times 10^{-2}$}&
\multicolumn{4}{c}{$Z=6.0\times 10^{-3}$}& \multicolumn{4}{c}{$Z=
10^{-3}$}} \startdata
$M/M_\odot$ & --- & 1.5 & 2.0 & 2.5 & 3.0 & 1.5 & 2.0 & 2.5 & 3.0 & 1.5 & 2.0 & 2.5 & 3.0 \\
\cline{1-14}
$^{86}$Kr/$^{84}$Kr   & 0.30 & 0.27 & 0.21  & 0.18 & 0.20  & 0.37  & 0.36  & 0.43  & 0.54  & 0.46  & 0.66  & 0.91  & 1.09  \\
$^{87}$Rb/$^{85}$Rb   & 0.38 & 0.38 & 0.30  & 0.27 & 0.30  & 0.53  & 0.57  & 0.85  & 0.89  & 0.81  & 1.29  & 1.61  & 1.79  \\
$^{96}$Zr/$^{94}$Zr   & 0.16 & 0.08 & 0.03  & 0.02 & 0.03  & 0.10  & 0.07  & 0.13  & 0.17  & 0.19  & 0.43  & 0.84  & 1.18  \\
$^{134}$Xe/$^{132}$Xe & 0.36 & 0.29 & 0.14  & 0.11 & 0.13  & 0.08  & 0.06  & 0.09  & 0.18  & 0.15  & 0.46  & 0.90  & 0.79  \\
$^{142}$Ce/$^{140}$Ce & 0.13 & 0.10 & 0.04  & 0.03 & 0.03  & 0.01  & 0.01  & 0.02  & 0.05  & 0.04  & 0.14  & 0.28  & 0.30  \\
\enddata
\label{tab10}
\end{deluxetable}

\begin{deluxetable}{lcccccccc}
\rotate \tablecolumns{9} \tablewidth{0pc} \tablecaption{Selected
physical and chemical quantities for a model with $M$=2 $M_\odot$
and $Z=6\times 10^{-3}$ (FRANEC Standard Model, FSM). Normalized
percentage variations as a function of different input parameters
are reported. Percentage values are defined as the difference
between Test cases and Standard Case, normalized to the Standard
one: (Test-FSM)/SM*100.}

\tablehead{Case &\colhead{$M^a_{\rm H}$} & \colhead{$\overline{\rm
M}$$_{\rm bol}^{C}$}& \colhead{$\tau_C$} &
\colhead{Yield$^a$($^{12}{\rm C}$)} &
\colhead{Yield$^a$($^{89}{\rm Y}$)} &
\colhead{Yield$^a$($^{139}{\rm La}$)} &
\colhead{Yield$^a$($^{208}{\rm Pb}$)} & \colhead{[hs/ls]}}
\startdata
 FSM             & 0.644    &  -4.98   & 1.38E+06  &  1.26E-02  & 1.00E-07  & 5.38E-08   & 1.50E-07 & 0.39    \\
$\alpha_1$       & +0.5\%   &  -13\%   &  +18\%    &   +39 \%   & +54\%     & +38\%      & +31\%    &  -15\%  \\
$\alpha_2$       & -1.6\%   &  +17\%   &  -22\%    & -36\%      & -47\%     & -40\%      & -42\%    &  +13\%  \\
$\dot M^{\rm AGB}_1$ & -3.1\%   & +20\%    & -28\%     & -30\%      & -29\%     & -24\%      & -21\%    &  7\%    \\
$\dot M^{\rm AGB}_2$ & +2.1\%   & -17\%    & +21\%     & 32\%       & 30\%      &  24\%      &  17\%    &  -5\%   \\
$\beta_1$        & -1.2\%   & +2\%     &   6\%     &   16\%     & -66\%     & -79\%      & -93\%    &   -53\% \\
$\beta_2$        & +1.0\%   & -9\%     & -16\%     & -19\%      &  -84\%    & -87\%      & -93\%    &  -36\%  \\
\cline{1-9} \\
\multicolumn{9}{l}{$^a $ In $M_\odot$ units} \\
\enddata \label{tab11}
\end{deluxetable}

\begin{deluxetable}{lccc}
\tablecolumns{4} \tablewidth{0pc} \tablecaption{Initial abundances
and net stellar yields relative to a 3 $M_\odot$ model with Z=0.02
taken from our database (FRUITY) and from \citealt{kakka10} (K10).
Percentage differences are also shown.}

\tablehead{Isotope & \colhead{Yields$^a$ (FRUITY)}&
\colhead{Yields$^a$ (K10)}& \colhead{[(FRUITY/K10)-1]*100} }
\startdata
$^{1}$H     &  -9.01E-02 &   -6.95E-02 &      -29   \\
$^{4}$He    &  7.36E-02  &   5.13E-02  &      +43   \\
$^{12}$C    &  1.10E-02  &   1.20E-02  &      -9   \\
$^{13}$C    &  1.27E-04  &   1.10E-04  &      +16   \\
$^{14}$N    &  4.16E-03  &   3.09E-03  &      +35   \\
$^{15}$N    &  -5.53E-06 &   -5.03E-06 &      -10  \\
$^{16}$O    &  -1.21E-03 &   -9.02E-04 &      -34  \\
$^{17}$O    &  7.12E-05  &   4.68E-05  &      +52   \\
$^{18}$O    &  -1.54E-05 &   -1.37E-05 &      -12  \\
$^{19}$F    &  2.25E-06  &   4.13E-06  &      -46 \\
$^{20}$Ne   &  -2.85E-06 &   1.01E-05  &      -128 \\
$^{21}$Ne   &  7.49E-07  &   1.05E-06  &      -29 \\
$^{22}$Ne   &  2.03E-03  &   3.45E-03  &      -41  \\
$^{23}$Na   &  6.67E-05  &   9.28E-05  &      -28  \\
$^{24}$Mg   &  3.03E-05  &   1.36E-05  &      +123  \\
$^{25}$Mg   &  3.56E-06  &   2.17E-05  &      -84 \\
$^{26}$Mg   &  1.40E-05  &   2.58E-05  &      -46 \\
$^{27}$Al   &  2.58E-06  &   2.58E-06  &      -19  \\
$^{28}$Si   &  -2.78E-07 &  -2.78E-07  &      -212 \\
$^{29}$Si   &  2.96E-07  &   2.96E-07  &      -19  \\
$^{30}$Si   &  1.28E-06  &   1.28E-06  &      +10   \\
$^{31}$P    &  8.06E-07  &   8.06E-07  &      -50 \\
\cline{1-4} \\
\multicolumn{4}{l}{$^a $ In $M_\odot$ units} \\
\enddata \label{tab12}
\end{deluxetable}

\end{document}